\pdfoutput=1
\PassOptionsToPackage{hyperfigures=true,colorlinks=true,linkcolor=OliveGreen,citecolor=Blue}{hyperref}
\documentclass[journal,onecolumn,draftclsnofoot]{IEEEtran}
\usepackage[T2A,T1]{fontenc}
\usepackage{t1enc}
\usepackage[utf8]{inputenc}
\usepackage[russian,english]{babel}
\usepackage{graphicx}
\usepackage{mathrsfs}
\usepackage{comment}
\usepackage{csquotes}
\usepackage{balance}
\usepackage{epstopdf}
\usepackage{psfrag}
\usepackage{amsmath}
\usepackage{amsfonts}
\usepackage{amssymb}
\usepackage{dsfont}
\usepackage[dvipsnames]{xcolor}
\usepackage{stmaryrd}
\usepackage{color}
\usepackage{amsthm}
\usepackage{epstopdf}
\usepackage{enumerate}
\usepackage{units}
\usepackage[super]{nth}
\usepackage{nicefrac}
\usepackage{kbordermatrix}

\usepackage{enumitem}
\newlist{abbrv}{itemize}{1}
\setlist[abbrv,1]{label=,labelwidth=1in,align=parleft,itemsep=0.1\baselineskip,leftmargin=!}
\DeclareMathOperator{\I}{I}
\DeclareMathOperator{\J}{JSD}
\DeclareMathOperator{\HH}{H}
\DeclareMathOperator{\DD}{D}

\def\iC{{\cal C}}

\def\iE{{\cal E}}

\def\iP{{\cal P}}
\def\iR{{\cal R}}
\def\iS{{\cal S}}
\def\iT{{\cal T}}

\def\iV{{\cal V}}

\def\iX{{\cal X}}
\def\iY{{\cal Y}}
\def\iZ{{\cal Z}}

\newcommand{\sumfrac}[2]{\genfrac{}{}{0pt}{}{#1}{#2}}
\newcommand{\eps}{\varepsilon}

\newcommand{\irott}{\mathcal}

\newcommand{\vL}{\mathbf L}

\newcommand{\vX}{\mathbf X}
\newcommand{\vx}{\mathbf x}
\newcommand{\vY}{\mathbf Y}
\newcommand{\vy}{\mathbf y}

\newcommand{\vz}{\mathbf z}

\newcommand{\vS}{\mathbf S}

\newcommand{\vi}{\mathbf i}
\newcommand{\vj}{\mathbf j}

\newcommand{\vV}{\mathbf V}

\theoremstyle{plain}
\newtheorem{Thm}{Theorem}

\newtheorem{Lem}{Lemma}
\newtheorem{Cor}{Corollary}
\theoremstyle{definition}
\newtheorem{Def}{Definition}

\theoremstyle{remark}
\newtheorem{Rem}{Remark}
\title{Error Exponents for Asynchronous Multiple Access Channels. Controlled Asynchronism may Outperform Synchronism}
\author{Imre Csiszár,
	Lóránt Farkas,
        Tamás Kói
\thanks{This paper has been presented in part at ISIT 2015 Hawaii, ISIT 2016 Hong Kong and ISIT 2019 Paris.}
\thanks{Imre Csisz\'ar is Professor Emeritus of the Alfréd Rényi Institute of Mathematics, Budapest;
e-mail: csiszar.imre@renyi.hu}
\thanks{L\'or\'ant Farkas is Assistant Professor of the department
of Analysis, Institute of Mathematics, Budapest University of Technology and Economics,
e-mail: lfarkas@math.bme.hu}
\thanks{Tamás Kói is Assistant Professor of the department
of Stochastics, Institute of Mathematics, Budapest University of Technology and Economics,
e-mail: koitomi@math.bme.hu}
\thanks{This work has been supported by the National Research, Development and Innovation Office – NKFIH K120706 and KH129601.}}

\begin{document}

\maketitle
\begin{abstract}
Exponential error bounds achievable by universal coding and decoding are derived for frame-asynchronous discrete memoryless 
multiple access channels with two senders, via the method of subtypes, a refinement of the method of types. Maximum empirical multi-information decoding is employed. A key tool is an improved packing lemma, that overcomes the technical difficulty caused by codeword repetitions, via an induction based new argument. The asymptotic form of the bounds admits numerical evaluation. This demostrates that error exponents achievable by synchronous transmission (if possible) can be superseeded via controlled asynchronism, i.e.  a deliberate shift of the codewords.
\end{abstract}


\section{Introduction}

\IEEEPARstart{D}{iscrete} memoryless multiple access channels (MACs) with two senders will be referred to as synchronous MAC (SMAC) or asynchronous MAC (AMAC) according to the senders' codeword transmissions are frame-synchronous or not. Symbol synchronism is always assumed, its absence could be addressed only in a continuous time model beyond the scope of this paper, see Verdú \cite{async-GMAC}. Different AMAC models are possible according to the allowed kinds of delay, see e.g. Farkas and Kói \cite{AMAC}. In this paper the term AMAC refers to the model when any deterministic delay is allowed that may be unknown to the senders or chosen by them.

The capacity region for SMAC has been determined by Ahlswede \cite{Ahlswede} and Liao \cite{Liao}, and for AMAC with arbitrary unknown delay by Poltyrev \cite{Poltyrev} and Hui and Humblet \cite{Hui-Humblet}. Gaps in \cite{Poltyrev,Hui-Humblet} were filled in the book of El Gamal and Kim \cite{ElGamal} and independently in \cite{AMAC}. The capacity region in the asynchronous resp. synchronous case is equal to the union for all $P^X$, $P^Y$ of the pentagons $\iR(P^X,P^Y,W)$ defined in \eqref{DefPentagon}, where $W$ is the channel matrix, respectively the convex closure of this union. As the union of these pentagons is non-convex for some choices of $W$ (Bierbaum and Wallmeier \cite{pelda}), the capacity region of SMAC may be larger than that of AMAC.

The error probability of good codes of block-length $n$, with rate pair inside the capacity region, goes to 0 exponentially as $n\to \infty$. The best possible exponent, called reliability function (as a function of the rate pair) is unknown. For SMAC, lower bounds to the reliability function, i.e. achievable  error exponents, have been derived by several authors, see Nazari et al. \cite{Nazari} and references there. Upper bounds were given by 
Harotounian \cite{Har75} and improved by Nazari et al. \cite{NazariUpper}.

Error exponents for AMAC, as far as we know, were first given by two of the present authors, reported in ISIT contributions \cite{Hawaii,Hongkong,Paris}. Upper-bounds to the reliability for AMAC are not available in the literature, and will not be given here, either. 

This paper is a completed, full version presentation of the results in \cite{Hawaii,Hongkong,Paris}. The main features are:
\begin{enumerate}[label=(\roman*)]
 \item Error exponents achievable universally, i.e., with codebooks and decoder not depending on the channel matrix, are derived via a refinement of the method of types (see Csiszár and Körner \cite{Csiszar}, Csiszár \cite{methodoftypes}), introduced in \cite{Hawaii} as method of subtypes. The universal achievability of our error exponents gives rise to a side result about the capacity region of compound AMAC.
 \item The adopted model is tailored to communication practice. The technical assumption in \cite{Hawaii,Hongkong} that the senders may use multiple codebooks is  no longer needed. This improvement relies upon a new twist in random selection  proof technique, reported in \cite{Paris}. \label{EnUjPacking}
 \item Our error exponents admit numerical evaluation, at least in simple cases. \label{EnNumerical}
 \item \label{EnumCAMAC}A remarkable discovery is that controlled asynchronism may beat synchronism: when synchronization would be possible, a deliberate shift of codewords may admit to achieve a larger error exponent than the (unknown) largest one for synchronous transmission, i.e., for SMAC. Evidence for this has been reported in \cite{Hongkong}, and a proof in \cite{Paris}. The proof uses numerical evaluation of the exponents, demonstrating the relevance of \ref{EnNumerical}.
\end{enumerate}

The method of subtypes has also been applied to exponents for multiple codebooks of unequal block-length (Farkas and Kói \cite{DiffWordlength,DiffBlocklength}), and for sparse communication by the present authors, \cite{SparseExponent}. Furthermore, Farkas and Kói \cite{ContSuccMAC} have analyzed successive decoding for AMAC via the subtype technique,  and showed that combined with controlled asynchronism it provides an alternative to rate splitting  (see Grant et al. \cite{Urbanke}),  when synchronization would be possible.

One of the main technical contribution is the proof of Lemma \ref{LemPacking-basic} underlying the result \ref{EnUjPacking}. It overcomes a technical obstacle to random coding proofs when codeword repetitions may occur. Its idea might prove useful also to other problems. In particular, this is expected for trellis codes for single-user error exponents, suggested by a relationship to be pointed out of AMAC codes to trellis code for single/user channels.

The device of shifting codewords is known to have benefits of several kinds in multi user communications, see e.g. Hou et al. \cite{HouSmeePfisterTomasin2006}, Gollakota and Katabi \cite{gollakota2008zigzag} or Emoto and Nazaki \cite{EmotoNozaki}. Result \ref{EnumCAMAC} identifies a new one, precisely formulated and proved within the adopted model. 

Finally, we emphasize that this paper focuses on theoretical results about potential capabilities of communication systems. As rather common in Shannon theory, their engineering relevance consists in giving insights, but further research is needed to turn them into results directly applicable to real systems.

\section{Preliminaries} 
\subsection{Notation}

The set $\{1,2,\dots,k\}$ is denoted by $[k]$. Logarithms and exponentials are to the base 2 i.e., $\log x=\log_2 x$, $\exp x =2^x$. Polynomial factors will be denoted by $p_n$.

Random variables (RVs) are assumed to take values in finite sets. These sets, the RVs, and their possible values are typically denoted by calligraphic letters  and corresponding upper and lower case italics such as $\irott X,X,x$. Boldface letters always denote (finite) sequences.

Probability distributions on finite sets are denoted by $P$ or $V$, the set of all distribution on $\iX$ is $\iP(\iX)$. Notations $V^{X},V^{XY}$ etc. mean (joint) distributions of the indicated RVs, and $V^{Y|X}$ denotes conditional distribution. The notation $P^X$, $P^Y$ will be reserved for distinguished distributions on $\iX$, $\iY$, and $P^{XY}$, $P^{XYZ}$ for the distribution on $\iX \times \iY$ respectively $\iX \times \iY \times \iZ$ defined by
\begin{align}\label{DefPXYZ}
  P^{XY}(x,y)\triangleq P^{X}(x)P^{Y}(y),\qquad P^{XYZ}(x,y,z)\triangleq P^{XY}(x,y)W(z|x,y)
\end{align}
the latter assuming a given channel matrix $W:\iX \times \iY \to \iZ$.

The \emph{type} (empirical distribution) of a sequence $\vx \in \mathcal{X}^n$ is denoted by $P_{\vx}$, similarly the joint types of two or more sequences (of equal length) by $P_{(\vx,\vy)}$ etc. The subset of $\mathcal{P} (\mathcal{X})$, $\mathcal{P} (\iX \times \iY)$, etc consisting of all types or joint types of length-$n$ sequences are denoted by $\mathcal{P}^n (\mathcal{X})$, $\mathcal{P}^n (\iX \times \iY)$, etc. For $V \in \mathcal{P}^n (\mathcal{X})$ or $V \in \mathcal{P}^n (\iX \times \iY)$, etc., the \emph{type class} $T^n_{V}$ is the subset  of $\mathcal{X}^n$ or $\mathcal{X}^n\times \mathcal{Y}^n$ consisting of sequences of type $P_{\vx}=V$ or pairs of sequences of joint type $P_{(\vx,\vy)}=V$, etc. 

Distributions, in particular types, are frequently represented as (joint) distributions of dummy RVs, e.g. for $V\in \iP(\iX \times \iY)$ we write $V=V^{XY}$. This convention often simplifies notation, e.g. the marginals of $V$ are simply $V^X$ and $V^Y$.

The distribution $V^{XYZ}\in \iP(\iX\times \iY \times \iZ)$ with given marginal $V^{XY}$ and conditional distribution $V^{Z|XY}=W$ will be denoted by $V^{XY}\circ W$, formally 
\begin{align}
 (V^{XY}\circ W)(x,y,z)\triangleq V^{XY}(x,y)W(z|x,y). \label{EqWkorVDef}
\end{align}
With this notation, $P^{XYZ}$ in \eqref{DefPXYZ} equals $P^{XY}\circ W$.

Our notation of information measures for RVs always indicates their (joint) distribution that the information measure really depends on.
E.g. $\HH_V(Y|X)$ means conditional entropy when $V^{XY}=V$. In an extended usage of this notation, $V$ may be a distribution on a product space larger than $\iX \times \iY$, then the understanding is that $V^{XY}$ equals the marginal of $V$ on $\iX \times \iY$. 

In addition to standard information measures, also multi-information of $m\geq 2$ RVs in the sense of Watanabe \cite{watanabe} will be frequently employed. It is
defined by
\begin{align}
 \I_V(X_1\wedge X_2 \wedge \dots \wedge X_m)\triangleq&\sum_{i=1}^m
 \HH_V(X_i)-\HH_V(X_1,X_2,\dots,X_m).\label{multi-info}
\end{align}
Note that multi-information of $m=2$ RVs equals mutual information.

For certain frequently occurring information quantities we will also use the following brief notations:

\begin{align}
 \I^0_V\triangleq \I_V(X\wedge Y), \quad \I^1_V\triangleq \I_V(X\wedge YZ), \quad \I^2_V\triangleq \I_V(Y\wedge XZ), \quad \I^{12}_V\triangleq \I_V(X\wedge Y \wedge Z). \label{DefMultiInfRovidites}
\end{align}

Empirical information measures for deterministic sequences, denoted by $\hat \HH$, $\hat \I$ are defined as information measures for dummy RVs whose joint distribution equals the joint type of the given sequences. For example
\begin{align}
 \hat \I(\vx_1\wedge \dots \wedge \vx_m)\triangleq& \I_{P_{(\vx_!,\dots,\vx_m)}}(X_1\wedge \dots \wedge X_m)\\
 =& \sum_{i=1}^m\hat\HH(\vx_i)-\hat \HH(\vx_1,\dots,\vx_m). \notag
\end{align}

The $L_1$-distance or variation distance in $\iP(\iX)$ is
\begin{align}
 \|P-V\|\triangleq \sum_{x\in \iX}|P(x)-V(x)| \label{DefVariationalDistance}
\end{align}
(in the literature, the latter term is often used for $\frac{1}{2}\|P-V\|$). 

Finally, $\iR(P^X,P^Y,W)$ will denote the pentagon
\begin{align}
 (R_1,R_2)\in \mathbb R^2:\left\{ \begin{array}{rrr}
                    0 \leq R_1&\leq& \I_P^1\\
                    0 \leq R_2&\leq&\I_P^2\\
                    R_1+R_2&\leq& \I_P^{12}
                   \end{array}
 \right.\label{DefPentagon}
\end{align}
where $P=P^{XYZ}=P^{XY}\circ W$, see \eqref{DefPXYZ}. Since $\I_P^0=\I_P(X\wedge Y)=0$, here instead of $\I_P^1$, $\I_P^2$, $\I_P^{12}$ one could also write $\I_V(X\wedge Z|Y)$, $\I_V(Y\wedge Z|X)$, $\I_V(XY \wedge Z)$ as more common in the literature.

\subsection{The model}\label{SubSecModelDef}

A discrete memoryless MAC with two senders is defined by two (finite) input alphabets $\mathcal{X}, \mathcal{Y}$, a (finite) output alphabet $\mathcal{Z}$, and a stochastic matrix $W: \mathcal{X} \times \mathcal{Y} \rightarrow \mathcal{Z}$. For input sequences $\vx\in \iX^n$, $\vy\in \iY^n$, the probability of output sequence $\vz\in \iZ^n$ is
\begin{align}
 W^n(\vz|\vx,\vy)\triangleq\prod_{i=1}^n W(z_i|x_i,y_i).
\end{align}
The matrix $W$ may be unknown to the senders and the receiver.

\begin{Def}\label{DefCode}
 A constant composition AMAC code of blocklength $n$, with rate pair $(R_1,R_2)$, is given by codebooks $\iC_1=\{\vx(i),i\in [2^{nR_1}]\}$, $\iC_2=\{\vy(j),j\in [2^{nR_2}]\}$, synch-sequences $\vx(0),\vy(0)$, and an integer $K\geq 2$. Here $\vx(0)$ resp. $\vy(0)$ and the codewords $\vx(i)$, $\vy(j)$ are distinct sequences\footnote{This distinctness condition, while not essential, will simplify the presentation of proofs.} in $\iX^n$ and $\iY^n$, each of the same type $P^X\in \iP^n(\iX)$ resp. $P^Y\in \iP^n(\iY)$.
\end{Def}

Senders 1 and 2 transmit codewords from $\iC_1$ resp. $\iC_2$, inserting the synch-sequence $\vx(0)$ resp $\vy(0)$ after each $K-1$ consecutive codewords. As synch-sequences do not carry information, the \emph{effective transmission rates} are $R_1\left(1-\frac{1}{K}\right)$, $R_2\left(1-\frac{1}{K}\right)$. We note that $R_1$, $R_2$, $P^X$, $P^Y$ may depend on $n$. Typically, $K$ is chosen sufficiently large to make the effective rates reasonably close to the nominal rates $R_1, R_2$, but for convenience not depending on the blocklength $n$.


Asynchronism causes a delay $0\leq D \leq nK-1$ of the sync-sequences of sender 1  relative to those of sender 2. The delay between codewords is denoted by $d$ , it is determined by
\begin{align} \label{Defd}
 d \equiv D \quad \pmod{n} & \qquad 0\leq d \leq n-1
\end{align}
The delay $D$ (as well as $d$) is either unknown to the senders or is chosen by them. The latter is referred to as \emph{controlled asynchronism}. It is assumed that the receiver is able to locate the sync-sequences. 

\begin{figure}[hbt]
\begin{center}
 \includegraphics[width=0.9\textwidth]{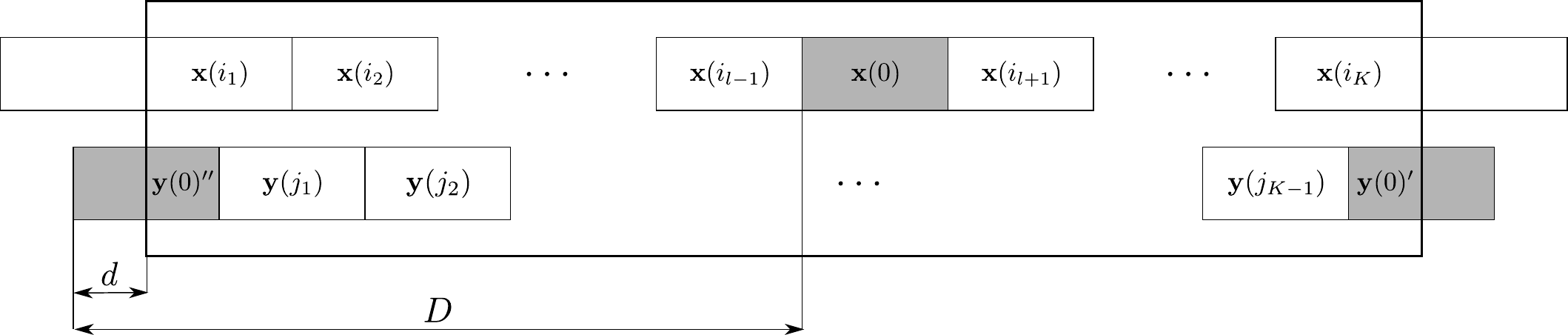}
 \caption{ The delays and the decoding window. The shaded blocks denote sync sequences. }\label{FigIntro}
\end{center}
\end{figure}

Decoding is performed in a window of length $nK$ shown in Figure \ref{FigIntro}. In this window, the the input sequence of sender 1 is the concatenation of $K-1$ codewords from $\iC_1$ and the synch-sequence $\vx(0)$, which is the $l$'th block where 
\begin{align}\label{DeflSzam}
    l \triangleq \frac{D-d}{n}+1.
\end{align}
The input sequence of sender 2 contains $K-1$ codewords from $\iC_2$, preceded and followed by complementary parts of the synch-sequence $\vy(0)$. 

\begin{Rem}
 Our adopting a model that involves synch-sequences is justified by communication practice. The receiver's ability to locate them is a reasonable assumption, since the probability of mislocating them is commonly negligible compared to decoding errors.
 For our purposes, the role of  synch-sequences is twofold. First, as the receiver can identify them, he/she knows the codewords boundaries. Second, the presence of synch-sequences admits the receiver to use a decoding window not containing split codewords, see Figure \ref{FigIntro}. Still, synch-sequences are not indispensable to achieve the AMAC error exponents in this paper, see Section \ref{SecConclusion}.
\end{Rem}

Most concepts below depend on the delay $D$. This dependence, supressed in the notation, is primarily through $d$ in \eqref{Defd}, the role of $l$ in \eqref{DeflSzam} will be less substantial. The concepts below are formulated under the assumption $d\ne0$, they need trivial modifications (actually simplifications) when $d = 0$. To save space, we temporarily exclude the case $d=0$, in effect, the synchronous case, until Section \ref{SecMainResult}. 

For an AMAC code in Definition \ref{DefCode}, the input sequences of senders 1 and 2 in the decoding window are
\begin{align}
 \vx(\vi)
 \triangleq&\vx(i_1)\dots \vx(i_{l-1})\vx(0)\vx(i_{l+1})\dots \vx(i_{K}), \quad \vi=i_1 \dots i_K \label{DefxMintIndexFv}
 \end{align}
 where $i_t\in[2^{nR_1}]$ for $t\in[K]\setminus \{l\}$ and $i_l=0$,
\begin{align}
 \vy(\vj)
 \triangleq&\vy(0)'' \vy(j_{1})\vy(j_{2})\dots \vy(j_{K-1})\vy(0)' \quad \vj=j_1\dots j_{K}\label{DefyMintIndexFv}
\end{align}
where $j_t\in[2^{nR_2}]$ for $t\in[K-1]$ and $j_K=0$, and $\vy'(0)$, $\vy''(0)$ are the length-$d$ initial resp. length-$(n-d)$ final parts of $\vy(0)$. In \eqref{DefxMintIndexFv}--\eqref{DefyMintIndexFv} $i_1,i_2,\dots,i_{l-1},i_{l+1},\dots,i_K$ and $j_1\dots,j_{K-1}$ represent the message $K-1$ tuples that senders 1 and 2 transmit in the considered decoding window. It is for technical convenience that the sequences $\vi$ and $\vj$ are taken to include also $i_l=0$ and $j_K=0$. In the sequel $\vi$ and $\vj$ (and similarly $\hat \vi$, $\hat \vj$) always denote sequences of length $K$ as above, in particular, the $l$-th element of $\vi$ (or $\hat \vi$) for $l$ in \eqref{DeflSzam} and $j_K$ (or $\hat j_K$) are always equal to 0.
\begin{Rem}\label{VirtualBlock}
 For convenience, the two parts of the sync sequence $\vy(0)$ in \eqref{DefyMintIndexFv} will be regarded as a single virtual block $\vy'(0)\vy''(0)=\vy(0)$ as if the decoding window were circular. Instances of some notations will formally refer to the 0'th or $K$'th block of the sequence \eqref{DefyMintIndexFv}, these will be interpreted to mean the virtual block.
\end{Rem}

The decoder employed in this paper will be specified later in Definition \ref{DefDekodolo}. Until then the decoder can be any mapping $\phi$ (depending on the delay $D$) that assigns estimates $\vx(\hat \vi)$,$\vy(\hat \vj)$ of the two input sequences to output sequences $\vz\in \iZ^{nK}$, or, equivalently, estimates $\phi(\vz)=(\hat \vi,\hat \vj)$ of $\vi,\vj$ in \eqref{DefxMintIndexFv},\eqref{DefyMintIndexFv} (satisfying the condition that $\hat i_l=0$ and $\hat j_K=0$).

For input sequences \eqref{DefxMintIndexFv},\eqref{DefyMintIndexFv} or equivalently for given $\vi,\vj$, and for output sequence $\vz\in\iZ^{nK}$, errorneous decoding means that $\phi(\vz)=(\hat \vi,\hat \vj)$ is not equal to $(\vi,\vj)$. As there are $2^{n(K-1)(R_1+R_2)}$ possible choices of $(\vi,\vj)$, the \emph{average probability of error} is
\begin{align}\label{ErrProb}
 P_e^D\triangleq 2^{-n(K-1)(R_1+R_2)} \sum_{\sumfrac{\vi,\vj,\hat \vi,\hat \vj}{(\hat \vi,\hat \vj)\neq(\vi,\vj)}}W^{nK}\left(\{\vz:\phi(\vz)=(\hat \vi,\hat \vj)\}\Big|\vx(\vi),\vy(\vj)\right).
\end{align}
Here, unlike in the previous notations, dependence on  the delay $D$ is not supressed, to emphatize that this dependence is a key issue of this paper. On the other hand, the dependence on the channel matrix $W$ is supressed, as also later in \eqref{DefErrProbPatt}.

Exponential upper bounds will be derived that hold for suitable AMAC codes even in universal sense: the AMAC code depends neither on the channel matrix $W$ nor on the delay $D$, and the decoder does not depend on $W$.

\begin{Rem}
 Assume that the senders' messages come from flows of independent random messages uniformly chosen from $[2^{nR_1}]$ resp. $[2^{nR_2}]$. Then the probability that not all members of these flows, transmitted in the given window, are decoded correctly is equal to $P_e^D$. Note that for sender 1 it depends on the delay $D$ which members of the infinite flow are transmitted in a particular window. Still, $P_e^D$ defined by \eqref{ErrProb} is closely related to another performance criterion, for individual members of these flows. Namely, for any coding-decoding system let $P_{e,ind}^D$  denote the supremum for all message indices $t$ and $i\in\{1,2\}$ of the probability of incorrect decoding of the $t$'th message of sender $i$. For our model, $P^D_{e,ind}$ equals $P_e^D$ up to constant factor: as one easily see,
 \begin{equation}
  P_{e,ind}^D \le P_e^D \le 2(K-1)  P_{e,ind}^D.
 \end{equation}
\end{Rem}

To bound the average probability of error \eqref{ErrProb}, the more refined problem of bounding \emph{error pattern} probabilities will be addressed. When sent sequences $(\vx(\vi),\vy(\vj))$ in \eqref{DefxMintIndexFv},\eqref{DefyMintIndexFv} are decoded as $(\vx(\hat \vi),\vy(\hat \vj))$ where 
\begin{align}\label{EPszam}
 \{t\in[K]:i_t\neq \hat i_t\}=L_1 \qquad \{t\in[K]:j_t\neq \hat j_t\}=L_2
\end{align}
we say that error pattern $(L_1,L_2)$ occurs, denoted by $(\vi,\hat \vi, \vj, \hat \vj)\in \iE\iP(L_1,L_2)$. The number $L=|L_1| + |L_2|$ of incorrectly decoded codewords is called the \emph{length} of this error pattern. Note that the blocks representing sync-sequences need not be considered in the definition of error pattern, as no error can occur there. So, the set $L_1$ resp. $L_2$ in \eqref{EPszam} never contains $l$ (given by \eqref{DeflSzam}) resp. $K$, and the largest possible length of an error pattern is $2K-2$.

The average probability of error pattern $(L_1,L_2)$ is
\begin{align}
 P_e^D(L_1,L_2)\triangleq 2^{-n(K-1)(R_1+R_2)} \sum_{(\vi,\vj,\hat \vi,\hat \vj)\in \iE\iP(L_1,L_2)}W^n\left(\{\vz:\phi(\vz)=(\hat \vi,\hat \vj)\}\Big|\vx(\vi),\vy(\vj)\right).\label{DefErrProbPatt}
\end{align}
Clearly,
\begin{align}
 P_e^D=\sum_{\sumfrac{L_1\subseteq [K]\setminus \{l\},L_2 \subseteq [K]\setminus \{K\}}{(L_1,L_2)\neq (\emptyset,\emptyset)}}P_e^D(L_1,L_2).\label{PeD-felosztasa-mintakra}
\end{align}
Formally also the error pattern $(\emptyset,\emptyset)$ is defined, it will be called improper error pattern, since $(\vi,\hat \vi, \vj, \hat \vj)\in \iE\iP(\emptyset,\emptyset)$ means $\hat \vi=\vi$, $\hat \vj=\vj$, i.e., no error.

\begin{figure}[hbt]
\begin{center}
\includegraphics[width=0.8\textwidth]{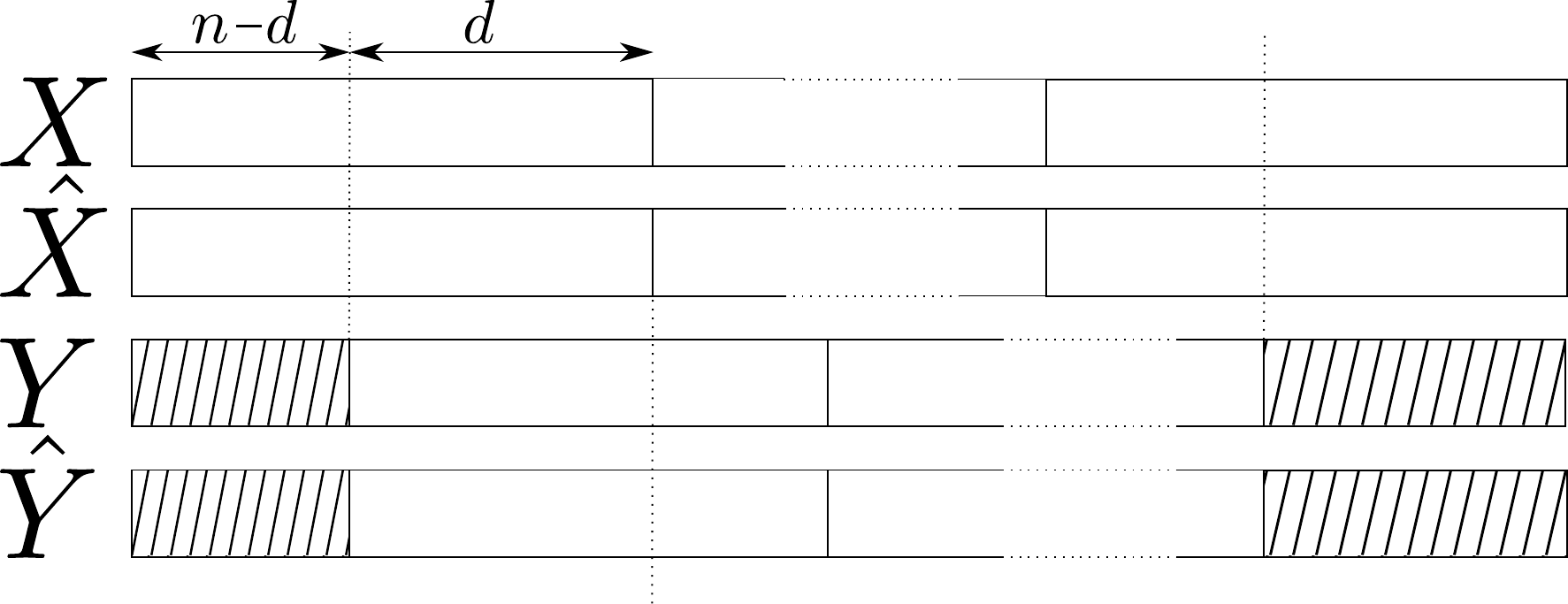}
\caption{ Array with four rows, partitioned into subblocks. The crosshatched parts denote synch-sequence. }\label{Fig4tuples}
\end{center}
\end{figure}

For bounding the probabilities \eqref{DefErrProbPatt}, an alternate characterization of error patterns will be useful. Arrange quadruples $\vx(\vi)$, $\vx(\hat \vi)$, $\vy(\vj)$, $\vy(\hat \vj)$ of potential input sequences and their estimates in an array with four rows, called rows $X$, $\hat X$, $Y$, $\hat Y$,  see Figure \ref{Fig4tuples}. The codeword boundaries partition the decoding window into $2K$ subintervals of length 
\begin{align}
 n_k=\left\{\begin{array}{cr}
            n-d & k \textnormal{ odd}\\
            d & k\textnormal{ even.}
           \end{array}
     \right.\label{DefSubBlockHossz}
\end{align}
Recall that the case $d=0$ has been excluded.

Accordingly, each block of the array in Figure \ref{Fig4tuples} is split into two subblocks giving rise to an array of subblocks. The $k$'th subblock in rows $X$ and $\hat X$ comes from the $t_1(k)$'th block in Figure \ref{Fig4tuples}, where 
\begin{align}
 t_1(k)=\left\{\begin{array}{cr}
            (k+1)/2 & k \textnormal{ odd}\\
            k/2 & k\textnormal{ even.}
           \end{array}
     \right.,
\end{align}
and in rows $Y$, $\hat Y$ from the $t_2(k)$'th block where 
\begin{align}
 t_2(k)=\left\{\begin{array}{cr}
            (k-1)/2 & k \textnormal{ odd}\\
            k/2 & k\textnormal{ even}
           \end{array}
     \right.
\end{align}
(for $k=1$, see Remark \ref{VirtualBlock}). Call $k$ an \emph{error index} of sender 1 or 2 if $\hat i_{t_1(k)}\neq i_{t_1(k)}$ or $\hat j_{t_2(k)}\neq j_{t_2(k)}$ respectively, i.e., $t_1(k)\in L_1$ or $t_2(k) \in L_2$, see \eqref{EPszam}. Let $S_1$, $S_2$ and $S_{12}$ denote the sets of those $k\in [2K]$ which are error indices for sender 1 but not 2, for sender 2 but not 1, and for both senders.  
As the error pattern $(L_1,L_2)$ of a quadruple $(\vi,\hat \vi,\vj,\hat \vj)$ is in a one-to-one correspondence with the triple $\vS=(S_1,S_2,S_{12})$ of disjoint subsets\footnote{But not all such triples correspond to error patterns} of $[2K]$, we will also speak of error pattern $\vS$, and use notation $(\vi,\hat \vi,\vj,\hat \vj)\in \iE\iP(\vS)$ as well as $P^D_e(\vS)$.  The set $S\triangleq S_1\cup S_2 \cup S_{12}$ is called the \emph{support} of error pattern $\vS$.

\begin{figure}[hbt]
\begin{center}
\includegraphics[width=0.6\textwidth]{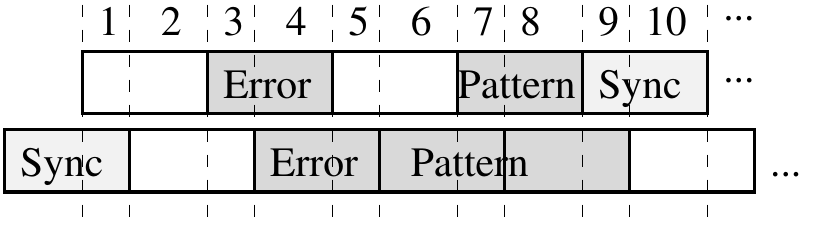}
\caption{ An example for subblocks and error pattern. Here $L_1=\{2,4\}$, $L_2=\{2,3,4\}$ and $S_1=\{3\}$, $S_2=\{5,6,9\}$, $S_{12}=\{4,7,8\}$. Thus, $\vS=(\{3\},\{5,6,9\},\{4,7,8\})$, $S=\{3,4,5,6,7,8,9,10\}$. }\label{FigSubblocks}
\end{center}
\end{figure}

A partial order among proper error patterns is defined,  letting $\vS'\prec \vS$ mean that $S_1'\subseteq S_1$, $S_2'\subseteq S_2$, $S_{12}'\subseteq S_{12}$. Error patterns $\vS$ with no subpattern $\vS'\prec \vS$, $\vS'\neq \vS$ will be called \emph{irreducible}. They are characterized by having support $S=\{k_1,k_1+1,\dots,k_1+L\}$, $L\in[2K-k_1]$, $k_1\in[2K-1]$,  and satisfying $S_{12}=\{k_1+1,\dots,k_1+L-1\}$. Here, as the notation suggests, $L=|S|-1$ equals the length $|L_1|+|L_2|$ of the irreducible error pattern $\vS=(L_1,L_2)$. For example the error pattern in Figure \ref{FigSubblocks}, of length 5, is not irreducible, although its support does consist of consecutive indices; It has two irreducible subpatterns of lenghts 2 and 3, with supports \{3,4,5\} and \{6,7,8,9\}.

As synch sequences can not contribute to error patterns, the irreducible error patterns possible for delay $D$ have length
\begin{equation} \label{eqLessl}
L \le 2 \max (l-1,K-l),
\end{equation}
with $l$ defined by (\ref{DeflSzam}).

\subsection{Technical Tools}\label{SubSecTechTools}

The error probabilities defined in the previous subsection will be bounded using an extension of the method of types \cite{Csiszar}, \cite{methodoftypes} to asynchronous models, introduced in \cite{Hawaii} as the method of subtypes. 

Below, the definition of subtypes is restricted, according to the needs of this paper, to sequences of length $nK$ partitioned into $2K$ subblocks of length defined by \eqref{DefSubBlockHossz}. These length-$nK$ sequences may, however, be arbitrary, not necessarily of form \eqref{DefxMintIndexFv} or \eqref{DefyMintIndexFv}.

\begin{Def}\label{DefSubtype}
Let sequences $\vx\in \iX^{nK}$, $\vy\in \iY^{nK}$, etc. be partitioned into $2K$ subblocks, 
denoted by $\vx_k$, $\vy_k$, of length $n_k$ in \eqref{DefSubBlockHossz}, $k\in[2K]$. The types of
these subblocks are called the \emph{subtypes} of $\vx$, $\vy$, etc. 
Similarly, for an $m$-tuple of length-$nK$ sequences, the joint type $V_k$ of their $k$'th subblocks is called the $k$'th subtype of this $m$-tuple. The set of all $m$-tuples of length-$nK$ sequences with given subtype sequence $\vV=(V_1,V_2,\dots,V_{2K})$ is denoted by $\iT_{\vV}=\iT_{V_1,V_2,\dots,V_{2K}}$\label{def-T_V}.
\end{Def}

For example, the subtypes of a triple $(\vx,\vy,\vz)\in \iX^{nK}\times \iY^{nK} \times \iZ^{nK}$ are the joint types $V_k=P_{(\vx_k,\vy_l,\vz_k)}$, $k\in[2K]$.Clearly 
\begin{align}
 |\iT_{\vV}|=\prod_{k=1}^{2K}|\iT_{V_k}^{n_k}|.\label{DefIrottTVvektor}
\end{align}

We emphasize that subtypes are defined relative to a given partition of $[2K]$, determined by the delay $D$ through \eqref{DefSubBlockHossz}. This dependence on $D$ is suppressed in the notation, as in case of other concepts.  

The next definition specifies the decoder employed in this paper. In the sequel, it will be assumed that the MMI decoder of Definition \ref{DefDekodolo} is used.
\begin{Def}\label{DefDekodolo}
 The maximal multi-information (MMI) decoder assigns to output sequence $\vz\in\iZ^{nK}$ that pair $(\vx(\hat \vi),\vy(\hat \vj))$ of potential input sequences $\vx(\vi),\vy(\vj)$ for which the weighted sum of empirical multi-informations
 \begin{align}
 \sum_{k=1}^{2K}n_k \hat \I(\vx_k(\vi)\wedge \vy_k(\vj)\wedge \vz_k)=\sum_{k=1}^{2K}n_k\I_{V_k}(X\wedge Y\wedge Z) \label{DefMultiinfodecoder}
\end{align}
 is maximal. Here $\vx_k(\vi)$, $\vy_k(\vj)$, $\vz_k$ are the $k$'th sub-blocks of the sequences \eqref{DefxMintIndexFv}, \eqref{DefyMintIndexFv} and $\vz$, according to Definition \ref{DefSubtype}, and $V_k$ is their joint type, i.e., $(V_1,\dots,V_{2K})$ is the subtype sequence of the triple $(\vx(\vi),\vy(\vj),\vz)$. If the maximizing $(\vi,\vj)$ is not unique, either one of them can be taken\footnote{Alternatively, in case of ties an error could be declared, this would lead to the same error bounds. For formal reasons, we prefer the decoder outputs to be always estimates of the sent messages.}.
\end{Def}
\begin{Rem}
    In the (temporarily excluded) case $d=0$, the terms in \eqref{DefMultiinfodecoder} of even index $k$ correspond to intervals of length 0 and are interpreted as 0, see \eqref{DefSubBlockHossz}. Thus, in that case, the sum actually has $K$ (rather than $2K$) terms, and $\vx_k(\vi)$, $\vy_k(\vj)$ represent full (rather than split) codewords. Then, maximization of the sum can be performed termwise. Further, maximizing $\hat \I(\vx_k(\vi) \wedge \vy_k(\vj) \wedge \vz_k)$ is equivalent to minimizing $\hat \HH(\vx_k(\vi), \vy_k(\vj)|\vz_k)$, since $\hat \HH(\vx_k(\vi))=\HH(P^X)$ and $\hat \HH(\vy_k(\vj))=\HH(P^Y)$ are constants. For SMAC, a decoder minimizing empirical conditional entropy has been used by Liu and Hughes in \cite{Hughes}. The decoder in Definition \ref{DefDekodolo} is its natural extension to AMAC.
    
\end{Rem}

\begin{figure}[hbt]
\begin{center}
\includegraphics[width=0.8\textwidth]{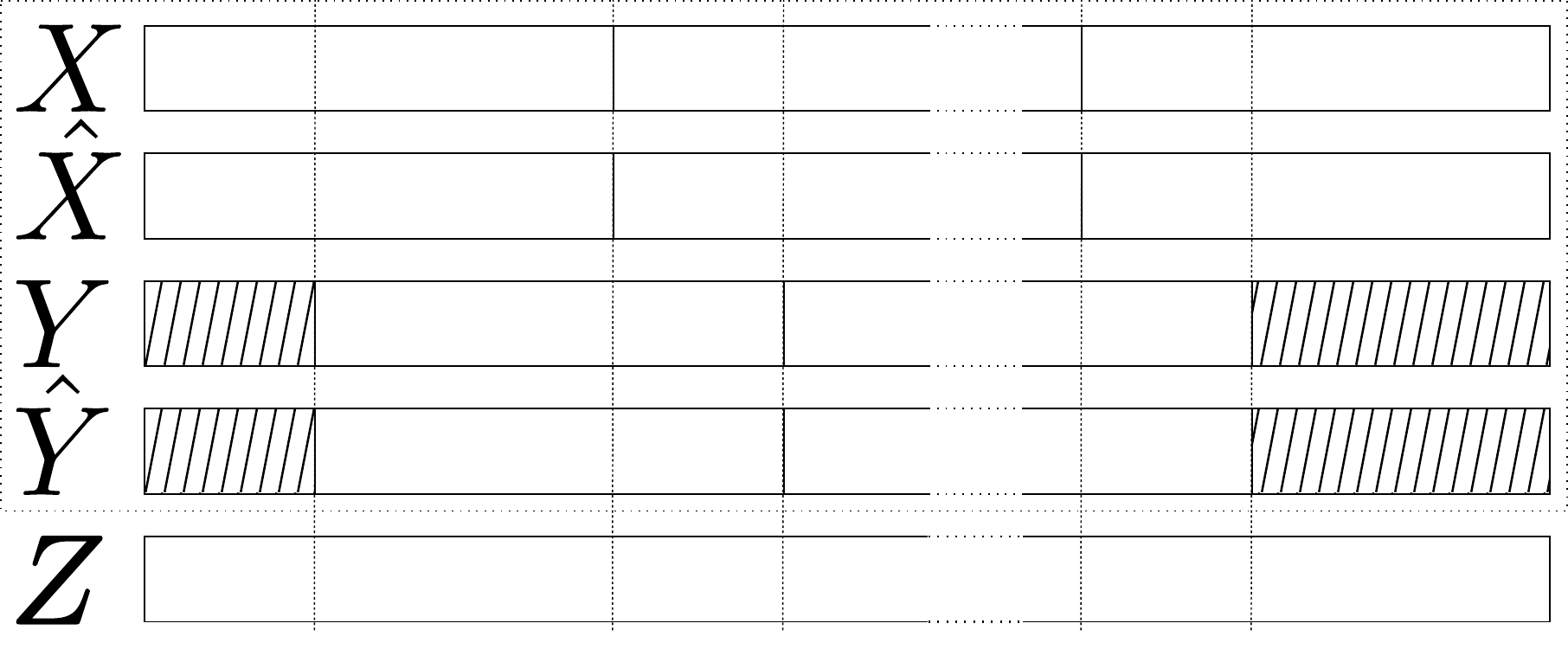}
\end{center}
\caption{ Array corresponding to quintuple $\left(\vx(\vi),\vx(\hat \vi),\vy(\vj),\vy(\hat \vj),\vz\right)$ }\label{Fig5tuples}
\end{figure}

A typical application of Definition \ref{DefSubtype} will be to quadruples $\left(\vx(\vi),\vx(\hat \vi),\vy(\vj),\vy(\hat \vj)\right)$ introduced in Subsection \ref{SubSecModelDef}. The $k$'th subtype $V_k$ of such a quadruple equals the joint type of the subblocks in the $k$'th column of the array of subblocks in Figure \ref{Fig4tuples}. We will also consider quintuples, with the channel output $\vz \in \iZ^{nK}$ added to the former sequences as a fifth one. This $\vz$ is also partitioned into $2K$ subblocks of length $n_k$, yielding a five-row array of subblocks, see Figure \ref{Fig5tuples}. The $k$'th subtype of this quintuple equals the joint type of the subblocks in the $k$'th column of the array with five rows. In these two cases, the subtypes will be denoted by $V_k^{X\hat XY\hat Y}$ respectively $V_k^{X\hat XY\hat YZ}$, conveniently indicating also that the former subtypes are marginals of the latter.

In calculations, the following standard facts (see e.g. \cite{Csiszar}) will be used, often without reference, for subtypes $V_k$ in the role of $V$ and $n_k$ in the role of $n$.
\begin{align}
 &|\iP^n(\iX)|\leq (n+1)^{|\iX|} \label{EqBasicfact1}\\
  &\frac{2^{n\HH_{V}(X)}}{(n+1)^{|\iX|}}\leq|T^n_{V}|\le 2^{n\HH_V(X)},\quad V\in \iP^n(\iX)  \label{EqBasicfact2-1}\\
   &|\{\vy\in \iY^n:P_{(\vx,\vy)}=V\}|\leq 2^{n\HH_V(Y|X)}, \quad \vx\in \iT^n_{V^X}\label{EqBasicfact4}\\
  &W^n(\vz|\vx,\vy)=\exp\{-n\left(\DD(V\|V^{XY}\circ W)+\HH_{V}(Z|XY)\right)\}, \quad V=P_{(\vx,\vy,\vz)}. \label{EqBasicfact3}
\end{align}

For the calculations it is inconvenient that the marginals of the subtypes $V_k^{X\hat XY\hat Y}$, i.e., the types of the subblocks of the array in Figure \ref{Fig4tuples} may be rather arbitrary in general, even though these subblocks are obtained by splitting blocks of fixed types $P^X$ or $P^{Y}$. This inconvenience will be overcome by chosing balanced codewords in the sense below. Then, the subblock types will be close to $P^X$ or $P^Y$, at least for not too short subblocks, see \eqref{EqBal4}. This is sufficient for our purposes, like ``typical sequences'' often adequately replace fixed type sequences.

\begin{Def}\label{DefDeltabalanced}
A sequence $\vx\in \iX^n$ of type $P_{\vx}=P$ will be called \emph{$\delta$-balanced} if for each $0<d<n$, the types $V_1$ and $V_2$ of its first $d$ and next $n-d$ symbols satisfy $\J^d_n(V_1,V_2)\leq \delta$, where
\begin{align}
 \J^d_n(V_1,V_2)\triangleq& \HH(P)-\frac{d}{n}\HH(V_1)-\frac{n-d}{n}\HH(V_2) \label{EqBal1}\\
 =&\frac{d}{n}\DD(V_1\|P) + \frac{n-d}{n}\DD(V_2\|P) \label{EqBal2}
 \end{align} 
 For $P\in\iP^n(\iX)$, the subset of $\iT_P^n$ consisting of $\delta$-balanced sequences will be denoted by $\iT^n_P(\delta)$.
 
 Note that in \eqref{EqBal1} 
 \begin{align}
  P=P_\vx=\frac{d}{n}V_1+\frac{n-d}{n}V_2.\label{EqBal3}
 \end{align}
\end{Def} 
The quantity $\J(\cdot)$ in \eqref{EqBal1} is known as Jensen-Shannon divergence. 
 
This concept of $\delta$-balanced sequences is of interest when $\delta$ is small. Then splitting any $\vx\in \iT^n_P(\delta)$ into two subblocks arbitrarily, their types $V_1$ and $V_2$ are close to $P$ (in $L^1$ distance), with the possible exception of short subblocks. Indeed $\J_n^d(V_1,V_2)\leq \delta$ implies by \eqref{EqBal2} and Pinsker inequality that
\begin{align}
 \|V_1-P\|<C\sqrt{\DD(V_1\|P)} \leq C\sqrt{\frac{n\delta}{d}}\label{EqBal4}
\end{align}
and similarly for $V_2$, with $d$ replaced by $n-d$, where $C=\sqrt{2\ln 2}$

The following key result of this subsection shows that for large $n$ a large fraction of any type class consists of $\delta$-balanced sequences, with $\delta$ admitted to go to 0 (sufficiently slowly) as $n\to \infty$. The lower bound in Lemma \ref{LemKicsiFnagy} on $\delta$ is crude but its order of magnitude $\frac{\log n }{n}$ is likely the best possible. The assertion of Lemma \ref{LemKicsiFnagy} holds for all $n\geq 2$ but it is trivial if $3|\iX|\log n \geq n\HH(P)$. Then the assumption implies $\delta\geq \HH(P)$ in which case all $\vx\in \iT_P^n$ are trivially $\delta$-balanced
\begin{Lem}[$\delta$-expurgating]\label{LemKicsiFnagy}
For each $P\in \iP^n(\iX)$ and $\delta\geq\frac{3|\iX|\log n}{n}$,
\begin{align}
|\iT^n_P(\delta)|\geq \frac{1}{2}|\iT^n_P|.\label{EqBal5} 
\end{align}
\end{Lem}
\begin{IEEEproof}
 Calculate the number of sequences $\vx\in \iT^n_P\setminus \iT^n_P(\delta)$. For some $d$, the types $V_1\in \iP^d(\iX)$ and $V_2\in \iP^{n-d}(\iX)$ of the subblocks $x_1x_2\dots x_d$ and $x_{d+1}x_{d+2}\dots x_n$ of such $\vx$ satisfy $\J_n^d(V_1,V_2)>\delta$. For fixed $d$ and $V_1$, $V_2$ there are
 \begin{align}
  \left| \iT^d_{V_1}\right|\cdot\left| \iT^{n-d}_{V_2}\right|\leq \exp\{d\HH(V_1)+(n-d)\HH(V_2)\}<\exp\{n(\HH(P)-\delta)\}
 \end{align} 
sequences with the latter property, due to \eqref{EqBasicfact2-1} and \eqref{EqBal1}. The number of admissible triples $(d,V_1,V_2)$ is less than
\begin{align}
 \sum_{d=1}^{n-1}\left|\iP^{d}(\iX) \right|<n^{|\iX|+1}
\end{align}
since $V_1$ determines $V_2$ by \eqref{EqBal3}. It follows that
\begin{align}
\left| \iT^n_P\setminus \iT^n_P(\delta) \right| < n^{|\iX|+1}\exp\{n(\HH(P)-\delta)\}.\label{EqBal6}
\end{align}
Simple algebra shows that for $\delta$ as in the Lemma the right hand side of \eqref{EqBal6} is less than $\frac{1}{2}(n+1)^{-|\iX|}\exp\{n\HH(P)\}$ thus less than $\frac{1}{2}|\iT^n_P|$ by \eqref{EqBasicfact2-1}.
\end{IEEEproof}

In the sequel, we denote 
\begin{align}
    \delta_n\triangleq \frac{3\log n}{n}\max(|\iX|,|\iY|).\label{EqDeltan}
\end{align}
Then, by Lemma \ref{LemKicsiFnagy}, \eqref{EqBal5} holds with $\delta=\delta_n$ for each $P^X\in\iP^n(\iX), P^Y\in\iP^n(\iY)$.

\section{Packing Lemma and Intermediate Form of the Exponential Error Bound}\label{sec-exponents}

The simplest random coding approach to exponential error bounds, both for single-user and multi-user channels, is to bound the error for random codes and conclude that then some deterministic code also meets this bound. That deterministic code, however, may be channel dependent. Universally achievable error bounds are commonly derived via so-called packing lemmas, that establish the existence of codes with ``good'' statistical properties. This existence proof typically employs random selection, but no matter how the existence is proven, any code with these ``good'' statistical properties does give the required error exponents simultaneously for all channels. 

To the AMAC error exponents in this paper, the following packing lemma will be the key. It asserts the existence of AMAC codes such that the number of quadruples $(\vi,\hat\vi,\vj,\hat\vj)\in \iE\iP(L_1,L_2)$ with $(\vx(\vi),\vx(\hat\vi),\vy(\vj),\vy(\hat\vj))$ having subtype sequence $\vV$ (i.e. belonging to $\iT_\vV$) is bounded above by a \emph{packing inequality} \eqref{EqPack-lem-ineq} , for all error patterns $(L_1,L_2)$ and subtype sequences $\vV=(V_1,V_2,\dots,V_{2K})$. We note that the possible subtype sequences of quadruples $(\vx(\vi),\vx(\hat\vi),\vy(\vj),\vy(\hat\vj))$ satisfy the constraints
\begin{align}
  \frac{n-d}{n}V^X_{2t-1}+\frac{d}{n}V^X_{2t}=P^X, \quad \frac{d}{n}V^Y_{2t}+\frac{n-d}{n}V^X_{2t+1}=P^Y, \quad  t \in [K]
\end{align}
(where $V^Y_{2K+1}$ is interpreted as $V^Y_1$), and those of quadruples with $(\vi,\hat\vi,\vj,\hat\vj)\in \iE\iP(L_1,L_2)=\iE\iP(\vS)$ also
\begin{align}
  &V^{X\hat{X}}_{k}(x,\hat x)=\mathds 1\{x=\hat x\}V_k^X(x) \text{ if }  k \notin S_1 \cup S_{12}\label{EqSubtypeProp1}\\
  &V^{Y\hat{Y}}_{k}(y,\hat y)=\mathds 1\{y=\hat y\}V_k^Y(y) \text{ if }  k \notin S_2 \cup S_{12}\label{EqSubtypeProp2}
\end{align}
where $\mathds 1\{x=\hat x\}$ resp. $\mathds 1\{y=\hat y\}$ equals 1 if $x=\hat x$ resp. $y=\hat y$, and 0 otherwise. Still, these constraints need not be included in the Lemma, since for ``impossible'' subtype sequences $\vV$ the packing inequalities trivially hold (the lhs of \eqref{EqPack-lem-ineq} equals 0).

\begin{Lem}\label{LemPacking-basic}
For each $K$, $n$, types $P^X\in\iP^n(\iX)$, $P^Y\in\iP^n(\iY)$, rates $R_1<\HH(P^X)-\delta_n,R_2<\HH(P^Y)-\delta_n$ and sets $\iT_1\subset \iT_{P^X}^n$, $\iT_2\subset \iT_{P^Y}^n$ of size not less than $\frac{|\iT^n_{P^X}|}{2}$ resp. $\frac{|\iT^n_{P^Y}|}{2}$ there exists an AMAC code with codewords and synch-sequences from $\iT_1$ resp. $\iT_2$ such that for each  $0 \le D \le Kn-1$, each error pattern $(L_1,L_2)$ (or $\vS=(S_1,S_2,S_{12})$) and each $\vV=(V_1, V_2,\dots, V_{2K})$ with $V_{k}\in \iP^{n_k}(\iX\times\iX\times\iY\times\iY)$, $k\in[2K]$ the following inequality holds:
 \begin{align}
  &\sum_{(\vi,\hat\vi,\vj,\hat\vj) \in \iE\iP(L_1,L_2)}\mathds{1}_{V_1, V_2,\dots, V_{2K}}\{\vx(\vi),\vx(\hat\vi),\vy(\vj),\vy(\hat\vj)\}\leq\notag\\
  &\leq p_n\exp\left\{ n(K-1)(R_1+R_2)\right\}\exp\left\{-\sum\limits_{k\notin S}n_k\left[\I_{V_k}(X \wedge  Y)\right]-\sum\limits_{k \in S_1}n_k\left[\I_{V_k}(\hat X \wedge X\wedge Y)-R_1\right]\right\}\notag\\ 
  &\cdot\exp\left\{-\sum\limits_{k \in S_2} n_k\left[\I_{V_k}(\hat Y \wedge X\wedge Y)-R_2\right]- \sum\limits_{k \in S_{12}} n_k\left[\I_{V_k}(\hat X \wedge \hat Y \wedge X\wedge Y)-R_1-R_2\right]\right\}\label{EqPack-lem-ineq}
 \end{align}
where $\delta_n$ is defined by \eqref{EqDeltan}. $\mathds{1}_{V_1, V_2,\dots,V_{2K}}\{\cdot\}$ denotes the indicator function of $\iT_\vV=\iT_{V_1, V_2,\dots,V_{2K}}$, the multi-information is defined by \eqref{multi-info} and $p_n$ is a polynomial in $n$ that depends only on $|\iX|$, $|\iY|$ and $K$.
\end{Lem}
In Lemma \ref{LemPacking-basic}, the improper error pattern $(L_1,L_2)=(\emptyset,\emptyset)$ is not excluded. The bound \eqref{EqPack-lem-ineq} for that specific case gives
  \begin{align}
    \sum_{\vi,\vj}\mathds{1}_{\tilde V_1,\dots,\tilde V_{2K}}(\vx(\vi),\vy(\vj))\leq p_n \exp\left\{n(K-1)(R_1+R_2)\right\}\exp\left\{-\sum_{k\in[2K]}n_k \I_{\tilde V_k}(X\wedge Y)\right\}\label{EqPack-lem-ineq-empty-patt}
  \end{align}
for each subtype sequence $(\tilde V_1,\dots,\tilde V_{2K})$ with $\tilde V_k\in \iP^{n_k}(\iX\times \iY)$, $k\in[2K]$, where the summation is for all pairs of message sequences $(\vi,\vj)$. Indeed, the lhs of \eqref{EqPack-lem-ineq-empty-patt} is equal to that of \eqref{EqPack-lem-ineq} for $(L_1,L_2)=(\emptyset,\emptyset)$ and $(V_1,\dots,V_{2K})$ defined by $V_k(x,\hat x,y,\hat y)=\tilde V_k(x,y)\mathds{1}\{x=\hat x,y=\hat y\}$.

The proof of this Lemma represents a major technical contribution of this paper. As it is rather lengthy, it will be given in Appendix \ref{Appendixpacking}. A weaker version, which is easy to prove, appears (in essence) in \cite{Hawaii}, where multiple codebooks are admitted.

In Theorem \ref{ThmHawaiiTovabbfejlesztve} below the following notations will be used. Recalling \eqref{DefMultiInfRovidites} and \eqref{EqWkorVDef}, denote for any sequence $\vV$ of distributions $V_k \in \mathcal{P}(\mathcal{X} \times \mathcal{Y} \times \mathcal{Z})$, $k \in [2K]$ (not necessarily types), any $\alpha\in [0,1]$ and error pattern $(L_1,L_2)$ or $\vS=(S_1,S_2,S_{12})$
 \begin{align}
  &E_\vV^\alpha(L_1,L_2)=E_\vV^\alpha(\vS)\triangleq \label{EVD}\\
  &\triangleq\sum_{k\in S}e_k \left[ \DD(V_k\|V_k^{XY}\circ W)+I_{V_k}^0 \right] +\left|\sum\limits_{k \in S_1}e_k (\I_{V_k}^1-R_1)+\sum\limits_{k \in S_2}e_k (\I_{V_k}^2-R_2)+\sum\limits_{k \in S_{12}}e_k (\I_{V_k}^{12}-R_1-R_2)\right|^+,\notag
 \end{align}
where 
 \begin{align}
  e_k\triangleq\left\{\begin{array}{ccr}
              1-\alpha & & k \textnormal{ odd} \\
              \alpha && k \textnormal{ even}
             \end{array}
 \right. . \label{Defek}
 \end{align}
Note that 
\begin{align} 
 e_k=\frac{n_k}{n}\textnormal{ if }\alpha=\frac{d}{n} \label{DefEk}.
\end{align}

Let $\iP_{n,k}$ be the collection of all distribution $V=V^{XYZ}\in\iP(\iX\times\iY\times\iZ)$ such that
\begin{align}
    &e_k\DD(V^X\| P^X)\leq \delta_n, \quad e_k\DD(V^Y\| P^Y)\leq \delta_n \label{EqMainThmJensenHelyett}
\end{align}
and set
\begin{align}
 E_n^\alpha(\vS)\triangleq&\min_{\vV:V_k \in \iP_{n,k}, k\in S} E_\vV^\alpha(\vS)  \label{DefEnAlphaS}.
\end{align}

\begin{Thm}[Intermediate form of the exponent]\label{ThmHawaiiTovabbfejlesztve}
  For all $K$, $n$, types $P^X\in \iP^n(\iX)$, $P^Y\in \iP^n(\iY)$, and rates $R_1\leq \HH(P^X)-\delta_n$, $R_2\leq \HH(P^Y)-\delta_n$ there exists an AMAC code as in Definition \ref{DefCode} such that for all channel matrix $W$, delay $D$, and proper error pattern $(L_1,L_2)$ or $\vS$ 
 \begin{equation} \label{fofobecsles-2}
  P_e^D(L_1,L_2)=P_e^D(\vS)\leq p_n\exp\left\{ -nE_n^\alpha(\vS)  \right\}, \quad \alpha=\frac{d}{n},
 \end{equation}
 where $p_n$ is a polynomial in $n$ that depends only on $|\iX|$, $|\iY|$, $|\iZ|$ and $K$. Consequently
 \begin{align}
   P_e^D \leq p'_n\exp\left\{ -n \min\limits_{\vS}E_n^\alpha(\vS) \right\}, \quad \alpha=\frac{d}{n}.\label{EqMainThmMasodik}
 \end{align}
where $p'_n=2^{2K-2}p_n$, and the minimum is over all proper error patterns $\vS$.
 
\end{Thm}
Of course the bound \eqref{fofobecsles-2} is of interest only when the exponent is positive, this issue will be addressed in Section \ref{SecMainResult}.
\begin{IEEEproof}
 We will show that each AMAC code that has the properties in Lemma \ref{LemPacking-basic}, with the choice $\iT_1=\iT^n_{P^X}(\delta_n)$, $\iT_2=\iT^n_{P^Y}(\delta_n)$, satisfies the assertions of Theorem \ref{ThmHawaiiTovabbfejlesztve}. The mentioned choice is permissible, due to Lemma \ref{LemKicsiFnagy} and \eqref{EqDeltan}.
 
 Fix such an AMAC code, fix also the delay $D$ and a proper error pattern $\vS=(S_1,S_2,S_{12})$. Denote by $\iV^{*}$ the collection of all subtype sequences $\vV=(V_1,V_2,\dots,V_{2K})$ of quintuples $(\vx(\vi),\vx(\hat \vi),\vy(\vj),\vy(\hat \vj),\vz)$ such that $(\vi,\hat\vi,\vj,\hat\vj)\in \iE\iP(\vS)$ and $\vz\in \iZ^{nK}$ is a channel output sequence that gives rise to decoder output $\phi(\vz)=(\hat\vi,\hat \vj)$. Then the components $V_k=V_k^{X\hat XY\hat YZ}$ of any $\vV\in \iV^{*}$ satisfy \eqref{EqSubtypeProp1}--\eqref{EqSubtypeProp2} (since $(\vi,\hat\vi,\vj,\hat\vj)\in \iE\iP(\vS)$), as well as the constraints \eqref{EqMainThmJensenHelyett} with $e_k=\frac{n_k}{n}$ (since the codewords and sync-sequences are from $\iT^n_{P^X}(\delta_n)$ resp. $\iT^n_{P^Y}(\delta_n)$, see Definition \ref{DefDeltabalanced}), and 
\begin{align}
 &\sum_{k\in [2K]}n_k \I_{V_k}(X \wedge Y \wedge Z)\le \sum_{k\in [2K]}n_k \I_{V_k}(\hat{X} \wedge \hat{Y} \wedge Z).\label{EqMainThmProofDecfuncIneq}
\end{align}
 (since $\vz$ is decoded as $(\hat \vi,\hat \vj)\neq(\vi,\vj)$)
 
 From \eqref{DefErrProbPatt} and \eqref{EqBasicfact3},
\begin{align}
 P_e^D (\vS)\leq& \sum_{\vV \in \iV^{*}} \left(\frac{1}{2^{nR_1}2^{nR_2}}\right)^{K-1}\sum_{\vi}\sum_{\vj}\prod_{k \in [2K]}\exp\left\{-n_k[\DD(V_k^{XYZ}||V_k^{XY}\circ W) + \HH_{V_k}(Z|XY)]\right\} \notag\\
 & \Big| \{\mathbf{z} \in \mathcal{Z}^{nK}:(\vx(\vi),\vx(\hat\vi),\vy(\vj),\vy(\hat\vj),\vz) \in T_{V_1, V_2,\dots,V_{2K}} \notag\\ 
 &\text{ for some } \hat\vi \text{ and } \hat\vj \text{ with } (\vi,\hat\vi, \vj,\hat\vj) \in \iE\iP (\vS) \}\Big|. \label{EqMainThmProofHibabecsles}
\end{align}
 The size of the set in \eqref{EqMainThmProofHibabecsles} is bounded above using \eqref{EqBasicfact4} in two different ways. The first upper-bound is
\begin{equation} \label{EqUpperbound1}
\exp\left[\sum\limits_{k \in [2K]}\hspace{-2mm}n_k \HH_{V_k}(Z|X Y)\right]\mathds{1}_{\tilde V_1, \tilde V_2,\dots, \tilde V_{2K}}\{\vx(\vi),\vy(\vj)\}
\end{equation}
where $\tilde V_k=V_k^{XY}$. The second bound is
\begin{align}
  &\sum_{\sumfrac{(\hat\vi,\hat\vj):}{(\vi,\hat\vi,\vj,\hat\vj) \in \mathcal{EP}(\vS)}} \mathds{1}_{V_1', V_2',\dots, V_{2K}'}\{\vx(\vi),\vx(\hat\vi),\vy(\vj),\vy(\hat\vj)\} \label{EqUpperbound2}\\ 
  &\exp\left\{\sum\limits_{k \notin S}n_k \HH_{V_k}(Z|X Y)+\sum\limits_{k \in S_1}n_k \HH_{V_k}(Z|X\hat X Y)+\sum\limits_{k \in S_2}n_k\HH_{V_k}(Z|X Y \hat Y)+\sum\limits_{k \in S_{12}}n_k\HH_{V_k}(Z|X \hat X Y \hat Y) \right\} \notag.
 \end{align}
 where $V_k'=V_k^{X\hat X Y\hat Y}$
Substituting \eqref{EqUpperbound1} in \eqref{EqMainThmProofHibabecsles} and employing \eqref{EqPack-lem-ineq-empty-patt} gives
\begin{align}
P_e^D (\vS)\leq& p_n\sum_{\vV \in \iV^{*}} \left(\frac{1}{2^{nR_1}2^{nR_2}}\right)^{K-1}\cdot \exp\left\{n(K-1)(R_1+R_2)\right\}\notag\\ 
  &\exp\left(-\sum_{k \in [2K]}n_k\DD(V_k^{XYZ}||V_k^{XY}\circ W)-\sum\limits_{k\in [2K]}n_k\I_{V_k}(X \wedge  Y)\right) \label{EqMainThmProofHibabecsles1}
\end{align}
On the other hand, substituting \eqref{EqUpperbound2} in \eqref{EqMainThmProofHibabecsles} and employing \eqref{EqPack-lem-ineq} gives
\begin{align}
P_e^D (\vS)\leq& p_n\sum_{\vV \in \iV^{*}} \left(\frac{1}{2^{nR_1}2^{nR_2}}\right)^{K-1}\cdot \exp\left\{n(K-1)(R_1+R_2)\right\}\notag\\ 
  &\exp\left(-\sum_{k \in [2K]}n_k\DD(V_k^{XYZ}||V_k^{XY}\circ W)-\sum_{k\in [2K]} n_k\HH_{V_k}(Z|XY)\right) \notag\\
  &\exp\left(-\sum_{k\notin S} n_k\I_{V_k}(X \wedge  Y)-\sum\limits_{k \in S_1}n_k\left[\I_{V_k}(\hat X \wedge X\wedge Y)-R_1\right]-\right.\notag\\ 
  &-\hspace{-2mm}\sum\limits_{k \in S_2}\hspace{-1mm} n_k\hspace{-1mm}\left[\I_{V_k}(\hat Y \wedge X\wedge Y)-R_2\right]-\notag\\
  &-\hspace{-2mm} \sum\limits_{k \in S_{12}}\hspace{-1mm} n_k \hspace{-1mm}\left[\I_{V_k}(\hat X \wedge \hat Y \wedge X\wedge Y)-R_1-R_2\right]- \notag\\
  &+\sum\limits_{k \notin S}n_k \HH_{V_k}(Z|X Y)+\sum\limits_{k \in S_1}n_k \HH_{V_k}(Z|X\hat X Y)+\notag\\
  &\left.+\sum\limits_{k \in S_2}n_k\HH_{V_k}(Z|X Y \hat Y)+\sum\limits_{k \in S_{12}}n_k\HH_{V_k}(Z|X \hat X Y \hat Y)\right) \label{EqMainThmProofHiabbecsles2}
\end{align}

We will use the inequalities
\begin{align}
  &\HH_{V_k}(Z|X\hat X Y)-\HH_{V_k}(Z|XY)-\I_{V_k}(X \wedge \hat X \wedge Y)=\notag\\
  &=-\I_{V_k}(X\wedge Y)-\I_{V_k}(\hat X \wedge ZXY)\leq -\I_{V_k}(X\wedge Y)-\I_{V_k}(\hat X \wedge YZ) \quad \textnormal{for }k\in S_1 \label{EqMainThmProofIdentity1} \\  
  &\HH_{V_k}(Z|X Y \hat Y)-\HH_{V_k}(Z|XY)-\I_{V_k}(X \wedge Y \wedge \hat Y)=\notag\\
  &=-\I_{V_k}(X\wedge Y)-\I_{V_k}(\hat Y \wedge ZXY)\leq-\I_{V_k}(X\wedge Y) -\I_{V_k}(\hat Y \wedge XZ) \quad \textnormal{for }k\in S_2 \label{EqMainThmProofIdentity2} \\
  &\HH_{V_k}(Z|X \hat X Y \hat Y)-\HH_{V_k}(Z|XY)-\I_{V_k}(X \wedge \hat X \wedge Y \wedge \hat Y)=\notag\\
  &=-\I_{V_k}(X\wedge Y)-\I_{V_k}(\hat X \wedge \hat Y \wedge ZXY)\leq -\I_{V_k}(X\wedge Y)-\I_{V_k}(\hat X \wedge \hat Y \wedge Z), \quad \textnormal{for }k\in S_{12} \label{EqMainThmProofIdentity3}
\end{align}
The bound \eqref{EqMainThmProofHibabecsles1} gives 
\begin{align}
  P_e^D (\vS)\leq& p_n\sum_{\vV \in \iV^{*}} \exp\left(-\sum_{k \in [2K]}n_k\left(\DD(V^{XYZ}_{k}\|V_k^{XY}\circ W)+\I_{V_k}(X\wedge Y)\right)\right) \label{EqMainThmProofBoundWithoutPattern}
\end{align}
and \eqref{EqMainThmProofHiabbecsles2} gives via \eqref{EqMainThmProofIdentity1}--\eqref{EqMainThmProofIdentity3}
\begin{align}
P_e^D (\vS)\leq& p_n\sum_{\vV \in \iV^{*}} \exp\left(-\sum_{k \in [2K]}n_k\left(\DD(V^{XYZ}_{k}\|V_k^{XY}\circ W)+\I_{V_k}(X\wedge Y)\right)\right)\cdot \notag\\
  &\exp\left(-\sum\limits_{k \in S_1}n_k\left[\I_{V_k}(\hat X \wedge YZ)-R_1\right]-\right.\notag\\ 
  &-\hspace{-2mm}\sum\limits_{k \in S_2}\hspace{-1mm} n_k\hspace{-1mm}\left[\I_{V_k}(\hat Y \wedge XZ)-R_2\right]-\notag\\
  &-\hspace{-2mm} \left.\sum\limits_{k \in S_{12}}\hspace{-1mm} n_k \hspace{-1mm}\left[\I_{V_k}(\hat X \wedge \hat Y \wedge Z)-R_1-R_2\right]\right) \label{EqMainThmProofBoundWithPattern}
\end{align}

Combining \eqref{EqMainThmProofBoundWithoutPattern} and \eqref{EqMainThmProofBoundWithPattern} and bounding the sum by the largest term times the number of terms, we obtain
 \begin{align}
  P^D_e(\vS)) \leq&p_n'\exp\left\{-\min_{\vV\in \iV^{*}}\left[\sum\limits_{k\in [2K]}n_k\left(\DD(V^{XYZ}_{k}\|V_k^{XY}\circ W)+\I_{V_k}(X\wedge Y)\right)+\right.\right.\notag\\
  +&\left|\sum\limits_{k \in S_1}n_k (\I_{V_k}(\hat X \wedge YZ)-R_1)+\right.\notag\\
  &+\sum\limits_{k \in S_2}n_k (\I_{V_k}(\hat Y \wedge XZ)-R_2)+\notag\\
  &+\left.\left.\left.\sum\limits_{k \in S_{12}}n_k (\I_{V_k}(\hat X \wedge \hat Y \wedge Z)-R_1-R_2)\right|^{+}\right]\right\} \label{EqMainThmProofHibabecsles3}
 \end{align}
 where $p_n'=p_n|\iV^*|$. 
 
 Next, the inequality \eqref{EqMainThmProofDecfuncIneq} will be invoked. There, $\I_{V_K}(\hat X\wedge \hat Y \wedge Z)$ is equal to $\I_{V_k}(X \wedge Y \wedge Z)$ if $k\notin S$, to $\I_{V_K}(\hat X\wedge Y \wedge Z)$ if $k \notin S_1$ and to $\I_{V_K}(X\wedge \hat Y \wedge Z)$ if $k \notin S_2$, see \eqref{EqSubtypeProp1},\eqref{EqSubtypeProp2}. Using this and (for $k\in S_1$ and $k\in S_2$) the identity 
 \begin{equation} \label{EqDekodoloFeltetelAtrendez}
\I_{V_K}(X\wedge Y \wedge Z)=\I_{V_K}(X\wedge Y,Z)+\I_{V_K}(Y \wedge Z)=\I_{V_K}(Y\wedge XZ)+\I_{V_K}(X\wedge Z),
 \end{equation} \eqref{EqMainThmProofDecfuncIneq} reduces to
\begin{align}
 &\sum\limits_{k \in S_1}n_k \I_{V_k}(\hat X \wedge YZ)+\sum\limits_{k \in S_2}n_k \I_{V_k}(\hat Y \wedge XZ)\notag\\
 &+\sum\limits_{k \in S_{12}}n_k \I_{V_k}(\hat X \wedge \hat Y \wedge Z) \geq \sum\limits_{k \in S_1}n_k \I_{V_k}( X \wedge YZ)\notag\\
 &+\sum\limits_{k \in S_2}n_k \I_{V_k}( Y \wedge XZ)+\sum\limits_{k \in S_{12}}n_k \I_{V_k}( X \wedge Y \wedge Z).  \label{EqMainThmProofIdentityForDecodingFunction}
\end{align}
	Using \eqref{EqMainThmProofIdentityForDecodingFunction} and  \eqref{EqMainThmProofHibabecsles3},\eqref{DefEk} gives 
	\begin{align}
	  P^D_e(\vS)\leq& p_n' \exp\left\{-n\min_{\vV \in\iV^{*}}\left[\sum\limits_{k\in [2K]}e_k\left[\DD(V^{XYZ}_{k}\|V_k^{XY}\circ W)+\I_{V_k}(X\wedge Y)\right]+\right.\right.\notag\\
	  &+\left|\sum\limits_{k \in S_1}e_k (\I_{V_k}(X \wedge YZ)-R_1)+\hspace{-2mm}\sum\limits_{k \in S_2} \hspace{-1mm}e_k (\I_{V_k}(Y \wedge XY)-R_2)\right.\notag\\
	  &\left.\left.\left.+\sum\limits_{k \in S_{12}}e_k (\I_{V_k}(X \wedge Y \wedge Z)-R_1-R_2)\right|^{+}\right]\right\}.\label{EqMainThmProofFofobecsles}
	\end{align}
	
	The expression to be minimized in \eqref{EqMainThmProofFofobecsles} depends only on the marginals $V_k^{XYZ}$ of the components $V_k$ of the subtype sequence $\vV \in \iV^*$. Hence, now letting $V_k$ denote distributions in $\iP^{n_k}(\iX \times \iY \times \iZ)$ (rather than in $\iP^{n_k}(\iX \times \iX \times \iY \times \iY \times \iZ)$ as before), the bound \eqref{EqMainThmProofFofobecsles} holds with $V_k^{XYZ}$ replaced by $V_k$ and minimizing over all sequences $\vV=(V_1,V_2,\dots,V_{2k})$ with $V_k\in \iP^{n_k}(\iX \times \iY \times \iZ)$ satisfying the constraints \eqref{EqMainThmJensenHelyett}. It holds even more if the minimum is taken over $2K$-tuples of any distributions $V_k\in\iP(\iX \times \iY \times \iZ)$ that satisfy \eqref{EqMainThmJensenHelyett}. When $\vV=(V_1,V_2,\dots,V_{2K})$  attains the latter minimum then $\DD(V_k^{XYZ}\|V_k^{XY}\circ W)+\I_{V_k}(X\wedge Y)=0$ for $k\notin S$. Thus the minimum in question is equal to the minimum of $E_\vV^\alpha(\vS)$ defined by \eqref{EVD}, i.e., to $E_n^\alpha(\vS)$. 
	This completes the proof of \eqref{fofobecsles-2}, since $p'_n$ above, though it does depend on $D$, is clearly bounded above by a polynomial that does not. Finally, \eqref{EqMainThmMasodik} obviously follows from \eqref{fofobecsles-2}.
\end{IEEEproof}

\begin{Thm}[Strengthening of Theorem \ref{ThmHawaiiTovabbfejlesztve}]\label{ThmWeakMonotonicity}
	There exist AMAC codes with the properties in Theorem \ref{ThmHawaiiTovabbfejlesztve} that meet additional bounds on the error pattern probabilities $P_e^D(\vS)$, namely
	\begin{equation} \label{fofobecslesuj}
	P_e^D(L_1,L_2)=P_e^D(\vS)\leq p_n2^{-nE_n^{\alpha}(\vS')}
	\end{equation}
	for each $\vS'\prec \vS$. Consequently, the bound \eqref{EqMainThmMasodik} on $P_e^D$ remains valid when restricting minimization there to irreducible error patterns.
\end{Thm}
\begin{Rem}
  Theorem \ref{ThmWeakMonotonicity} would trivially follow from Theorem \ref{ThmHawaiiTovabbfejlesztve} if $\vS'\prec \vS$ always implied $E_n^\alpha(\vS')\leq E_n^\alpha(\vS)$. The validity of that implication remains open.
\end{Rem}
\begin{IEEEproof}
  The proof of Theorem \ref{ThmWeakMonotonicity} differs from that of Theorem 1 only in one detail, namely bounding the set size in \eqref{EqMainThmProofHibabecsles} also in other ways than there. A new observation we also need is that the subtype sequences $\vV=(V_1,\dots,V_{2K})\in \iV^*$ satisfy, in addition to the properties used in the proof of Theorem \ref{ThmHawaiiTovabbfejlesztve}, also
	\begin{equation} \label{EqProofThmWMFeltetelVre}
	  \sum_{k\in S'} n_k \I_{V_k}(X \wedge Y \wedge Z)\le \sum_{k\in S'} n_k \I_{V_k}(\hat{X} \wedge \hat{Y} \wedge Z).
	\end{equation}
  for each $\vS'\prec \vS$ ($S'$ denotes the support of $\vS'$). To verify \eqref{EqProofThmWMFeltetelVre}, recall that $\vV' \in \iV^*$ means that $\vV$ is the subtype sequence of some quintuple $(\vx(\vi),\vx(\hat \vi),\vy(\vj),\vy(\hat \vj),\vz)$ such that $(\vi,\hat\vi,\vj,\hat\vj)\in \iE\iP(\vS)$     and $\vz$ is decoded as $\phi(\vz)=(\hat\vi,\hat\vj)$. Thus \eqref{EqProofThmWMFeltetelVre} means that for such quintuples 
    \begin{equation} \label{EqProofThmWMLenyeg}
      \sum_{k \in S'} n_k \hat \I(\vx_k(\hat \vi)\wedge \vy_k(\hat \vj)\wedge \vz_k) \ge  \sum_{k \in S'} n_k \hat \I(\vx_k(\vi)\wedge \vy_k(\vj)\wedge \vz_k) 
    \end{equation}
  If \eqref{EqProofThmWMLenyeg} failed for some $\vS'\prec\vS$, say $\vS'=(L_1',L_2')$, then changing the components $\hat i_t, t\in L_1'$ and $\hat j_t, t\in L_2'$ of $\hat \vi$ and $\hat \vj$ to $i_t$ respectively $j_t$ would give rise to a pair $(\hat\vi',\hat\vj')$ that outperforms $(\hat\vi,\hat\vj)$ in terms of the MMI decoding criterion, contradicting $\phi(\vz)=(\hat \vi,\hat\vj)$.
  
  Note that \eqref{EqProofThmWMFeltetelVre} is equivalent to 
  \begin{align}
    &\sum\limits_{k \in S'_1}n_k \I_{V_k}(\hat X \wedge YZ)+\sum\limits_{k \in S'_2}n_k \I_{V_k}(\hat Y \wedge XZ)\notag\\
    &+\sum\limits_{k \in S'_{12}}n_k \I_{V_k}(\hat X \wedge \hat Y \wedge Z) \geq \sum\limits_{k \in S'_1}n_k \I_{V_k}( X \wedge YZ)\notag\\
    &+\sum\limits_{k \in S'_2}n_k \I_{V_k}( Y \wedge XZ)+\sum\limits_{k \in S'_{12}}n_k \I_{V_k}( X \wedge Y \wedge Z).  \label{EqProofThmWMId5Ujra}
  \end{align}
  by the same argument as \eqref{EqMainThmProofDecfuncIneq} reduces to \eqref{EqMainThmProofIdentityForDecodingFunction}.
  
  The new upper bound to the size of the set in \eqref{EqMainThmProofHibabecsles} is 
  \begin{align}
    &\sum_{\sumfrac{(\hat\vi,\hat\vj):}{(\vi,\hat\vi,\vj,\hat\vj) \in \mathcal{EP}(L'_1,L'_2)}} \mathds{1}_{V^{'}_1, V^{'}_2,\dots, V^{'}_{2K}}\{\vx(\vi),\vx(\hat\vi),\vy(\vj),\vy(\hat\vj)\} \label{EqProofThmWMUpperbound2}\\ 
    &\exp\left\{\sum\limits_{k \notin S'}n_k \HH_{V_k}(Z|X Y)+\sum\limits_{k \in S'_1}n_k \HH_{V_k}(Z|X\hat X Y)+\sum\limits_{k \in S'_2}n_k\HH_{V_k}(Z|X Y \hat Y)+\sum\limits_{k \in S'_{12}}n_k\HH_{V_k}(Z|X \hat X Y \hat Y) \right\} \notag
  \end{align}
  where $(V_1',\dots,V_{2K}')$ denotes the subtype sequence of quadruples defined by $V_k'=V_k^{X\hat XY\hat Y}$ if $k\in S'$ and $V_k'(x,\hat x,y,\hat y)=V_k^{XY}\mathds 1\{x=\hat x\}\mathds 1\{y=\hat y\}$ if $k\notin S'$. Applying this bound in \eqref{EqMainThmProofHibabecsles}, and the packing inequality \eqref{EqPack-lem-ineq} with $\vS'$ in the role of $\vS$ (as the codebooks were chosen according to Lemma \ref{LemPacking-basic}, \eqref{EqPack-lem-ineq} holds for all error patterns), we obtain the analogue of \eqref{EqMainThmProofHiabbecsles2} with $\vS$ replaced by $\vS'$ on the right hand side.
  
  Employing instead of \eqref{EqMainThmProofHiabbecsles2} its analogue as above, the proof of Theorem \ref{ThmWeakMonotonicity} is completed exactly as that of Theorem \ref{ThmHawaiiTovabbfejlesztve}, using in the step when \eqref{EqMainThmProofIdentityForDecodingFunction} has been used now its analogue \eqref{EqProofThmWMId5Ujra}.
\end{IEEEproof}

\section{Asymptotic form of the exponents}\label{SecSimplify}

The error exponents in Theorems \ref{ThmHawaiiTovabbfejlesztve} and \ref{ThmWeakMonotonicity}, though given by single letter expressions, are prohibitively complex for computation. Below, the  exponents will be simplified in the limit $n\to \infty$, arriving at a form suitable for numerical evaluation, at least for simple channels. We note that this does not necessarily provides an (approximate) evaluation of the exponents $E_n^\alpha(\vS)$ for blocklength $n$ in communication practice, see Remark \ref{RemNemjoElmeletGyakban} in Appendix \ref{Appendixfolyt}.

Recall that $E_n^\alpha(\vS)$ has been defined in \eqref{DefEnAlphaS} as the minimum of $E_\vV^\alpha(\vS)$ in \eqref{EVD} over those sequences $\vV=(V_1,V_2,\dots,V_{2K})$ that satisfy $V_k\in \iP_{n,k}$ for $k\in \iS$, i.e., for all $k$ $V_k$ fullfill the constraints
    \begin{align}
         &e_k\DD(V_k^X\| P^X)\leq \delta_n, \quad e_k\DD(V_k^Y\| P^Y)\leq \delta_n, \label{EqImreDel}
    \end{align}
where $\delta_n$ and $e_k$ are defined by \eqref{EqDeltan} and \eqref{Defek}. Let $E^\alpha(\vS)$ denote the minimum subject to the stronger constraints $V_k^X=P^X$, $V_k^Y=P^Y$ for each $k\in S$, i.e., 
    \begin{align}
        E^\alpha(\vS)\triangleq \min_{\vV:V_k\in \iP^*, k\in S} E^\alpha_\vV(S),  \label{EqEalphaS}
    \end{align}
where $\iP^*$ denotes the collection of all distributions $V\in \iP(\iX\times \iY \times \iZ)$ with $V^X=P^X$, $V^Y=P^Y$. 

In this section, formal properties of $E_n^\alpha(\vS)$ and $E^\alpha(\vS)$ will be established, for arbitrary $\alpha\in[0,1]$, $P^X\in \iP(\iX)$, $P^Y\in \iP(\iY)$. The convergence $E_n^\alpha(\vS)\to E^\alpha(\vS)$ is rather plausible. It will be essential that this is uniform is $\vS$, $\alpha$, $P^X$, $P^Y$, $R_1$, $R_2$ and $W$, which appears non-trivial.

\begin{Thm}\label{ThmExpEgyenletesFolyt}
    $E_n^\alpha(\vS)$ converges uniformly to $E^\alpha(\vS)$, i. e.,
    \begin{align}
        E_n^\alpha(\vS)\geq E^\alpha(\vS)-\gamma_n, \quad \gamma_n\to 0, \label{EqThmEgyFolytAlsoKorlat}
    \end{align}
    where $\gamma_n$ depends only on $|\iX|$, $|\iY|$, $|\iZ|$ and $K$.
\end{Thm}

 A proof will be given in Appendix \ref{Appendixfolyt}.

 \begin{Rem}
  The rather tedious proof of Theorem \ref{ThmExpEgyenletesFolyt} can be dispensed with when dealing with controlled asynchronous transmission. Namely, a minor modification of the proof of Theorems \ref{ThmHawaiiTovabbfejlesztve}  resp. \ref{ThmWeakMonotonicity} shows that for fixed $\alpha=\frac{d}{n}$ there exist AMAC codes that meet the bounds \eqref{fofobecsles-2} or \eqref{fofobecslesuj} with $E^\alpha(\vS)$ or $E^\alpha(\vS')$ rather than $E_n^\alpha(\vS)$ or $E_n^\alpha(\vS')$ in the exponent. The modification consists in using Lemma \ref{LemPacking-basic} with $\iT_1\triangleq \iT^{n-d}_{P^X}\times \iT^{d}_{P^X}$, $\iT_2\triangleq \iT^{d}_{P^Y}\times \iT^{n-d}_{P^Y}$ (assuming that the $n$-types $P^X$, $P^Y$ are also $d$-types and hence $(n-d)$-types, as well). Although $|\iT_1|\geq\frac{1}{2}|\iT^{n}_{P^X}|$ and $|\iT_2|\geq\frac{1}{2}|\iT^{n}_{P^Y}|$ do not hold for these sets, similar bounds with 2 replaced by a polynomial of $n$ that do hold are sufficient for the assertion of the Lemma. Selecting the codewords from these $\iT_1$, $\iT_2$, the subtype sequences that may occur have all marginals $V_k^X$ resp. $V_k^Y$ equal to $P^X$ resp. $P^Y$, rather than satisfy only \eqref{EqMainThmJensenHelyett}.
 \end{Rem}

Next, the form of $E^\alpha(\vS)$ will be simplified, showing that the minimum in \eqref{DefEnAlphaS} for sequences $\vV$ of distributions $V_k\in \iP^*$ is attained when $V_k$ depends only on whether the index $k$ belongs to $S_1$, $S_2$ or $S_{12}$. Thus, to evaluate $E^\alpha(\vS)$, minimization over triples of distributions in $\iP^*$ suffices.

\begin{Thm}[Simpler form of $E^\alpha(\vS)$] \label{ThmConvexityFinalform}
  For each proper error pattern $\vS$,
  \begin{align}
    E^\alpha(\vS)=& \min_{V_1,V_2, V_{12} \in \iP^*} \left[ \sum_{i\in\{1,2,12\}}\beta_i \DD(V_{i}\|P^{XYZ}) + \left|\sum_{i\in\{1,2,12\}}\beta_i (\I_{V_i}^i-R_i)\right|^{+}  \right], \label{EqEgyszeruExp}
  \end{align}
   where $P^{XYZ}$ is defined by \eqref{DefPXYZ}, $R_{12} \triangleq R_1+R_2$, and $\beta_i \triangleq \sum_{k\in S_i} e_k$, $i\in\{1,2,12\}$.
\end{Thm}
\begin{IEEEproof}
  $E_{\vV}^\alpha(\vS)$ defined in \eqref{EVD} depends only on the distributions $V_k$ in the sequence $\vV$ with index $k\in S$. When these $V_k$ are in $\iP^*$, i.e., have marginals $P^X$, $P^Y$ then
  \begin{align} \label{EqMainThmProofDivExchange}
    \DD(V^{XYZ}_{k}\|V_k^{XY}\circ W)+\I_{V_k}(X\wedge Y)&=\DD(V^{XYZ}_{k}\|P^{XYZ})-\DD( V_k^X||P^X)-\DD( V_k^Y||P^Y)\notag\\
    &=\DD(V^{XYZ}_{k}\|P^{XYZ}) 
  \end{align}
  This implies using \eqref{DefMultiInfRovidites} that
  \begin{align}
    E_{\vV}^\alpha(\vS)=&\sum\limits_{k\in S_1}e_k\DD(V_{k}\|P^{XYZ})+\sum\limits_{k\in S_2}e_k\DD(V_{k}\|P^{XYZ})+\sum\limits_{k\in S_{12}}e_k\DD(V_{k}\|P^{XYZ})\notag\\
    &+\left|\sum\limits_{k \in S_1}e_k (\HH(P^X)-\HH_{V_k}(X|YZ)-R_1)+\right.\notag\\
    &\quad+\sum\limits_{k \in S_2}e_k (\HH(P^Y)-\HH_{V_k}(Y|XZ)-R_2)+\\
    &\quad\left.+\hspace{-1mm}\sum\limits_{k \in S_{12}}\hspace{-2mm}e_k (\HH(P^X)+\HH(P^Y)-\HH_{V_k}(XY|Z)-R_1-R_2)\right|^+\notag
 \end{align}
 Since divergence is convex and conditional entropy is concave in $V_k$, it follows for each collection $V_k\in\iP^*, k\in S$ that 
 \begin{align}
  E_\vV^\alpha(\vS)\geq E_{\hat \vV}^\alpha(\vS)
 \end{align}
 where $\hat V_l=$ $\frac{\sum_{k\in S_i}e_k V_k}{\sum_{k \in S_i}e_k}$ if $l \in S_i$, $i\in\{1,2,12\}$. This proves \eqref{EqEgyszeruExp}.
\end{IEEEproof}

\section{Main Results} \label{SecMainResult}

For irreducible error patterns $\vS$, with support $S=\{k_0,k_0+1,\dots,k_0+L\}$, the coefficients $\beta_i, i \in \{1,2,12\}$ in Theorem \ref{ThmConvexityFinalform} are given by
\begin{enumerate}[label=\emph{\alph*})]
	\item odd $L$, odd $k_0$: $\beta_1=1$, $\beta_2=0$,$\beta_{12}=\frac{L-1}{2}$
	\item odd $L$, even $k_0$: $\beta_1=0$, $\beta_2=1$,$\beta_{12}=\frac{L-1}{2}$
	\item even $L$, odd $k_0$: $\beta_1=\beta_2=1-\alpha$, $\beta_{12}=\alpha + \frac{L}{2}-1$
	\item even $L$, even $k_0$: $\beta_1=\beta_2=\alpha$, $\beta_{12}=\frac{L}{2}-\alpha$.
\end{enumerate}  
The corresponding exponent in \eqref{EqEgyszeruExp} depends only on the length $L$ of the (irreducible) error pattern $\vS$ and on the parity of $k_0$, i.e., whether the pattern starts with an error for sender $1$ or sender $2$. Denote this exponent by $E^\alpha (L,j)$, i.e., 
\begin{align}
E^\alpha (L,j) \triangleq &\min_{V_1,V_2, V_{12} \in \iP^*} \left[ \sum_{i \in \{1,2,12\}} \beta_i \DD(V_{i}\|P^{XYZ}) + \left|\sum_{i \in \{1,2,12\}} \beta_i (\I_{V_i}^i -R_i)\right|^+ \right] \label{ELJ} 
\end{align}
where $\beta_1,\beta_2,\beta_{12}$ are given by (a) or (c) above if $j=1$, and by (b) or (d) if $j=2$, and $R_{12}\triangleq R_1+R_2$. Recall that $\I_V^1\triangleq \I_V(X\wedge YZ)$, $\I_V^2\triangleq \I_V(Y\wedge XZ)$, $\I_V^{12}\triangleq \I_V(X\wedge Y \wedge Z)$. Remarkably, $E^\alpha (L,j)$ does depend on $\alpha$ only when $L$ is even. Define further for integers $M \ge 2$,
\begin{equation}
  E^{\alpha,M} \triangleq \min_{L\in[M], j\in\{1,2\}} E^\alpha (L,j).
  \label{DefAsExp}
\end{equation}
These exponents are well defined for all $\alpha \in [0,1]$ and $P^X \in \iP(\iX)$, $P^Y \in \iP(\iY)$. Their dependence on $P^X$, $P^Y$, as well as on $R_1$, $R_2$ and the channel $W$, is suppressed in the notation.  Note that, while $E^\alpha(L,j)$ is defined by $(L,j)$ independently of $K$, its meaning as error exponent is limited to $L\in[2K-2]$, i.e., $K\geq \frac{L}{2}+1$.

From this point on, the case $d=0$ (in effect, the synchronous case) is no longer excluded. In that case, the only irreducible patterns are those that represent error in a single codeword position, for one sender or both. Formally, with the notation (\ref{EPszam}), these are the error patterns $(L_1,L_2)$ with one of the sets $L_1$, $L_2$ a singleton and the other empty, or $L_1$, $L_2$ both equal to the same singleton. The exponents corresponding to these error patterns are $E^{0}(1,1)$, $E^{0}(1,2)$ and $E^{0}(2,2)$, given by (\ref{ELJ}) with $\beta_1$ or $\beta_2$ or $\beta_{12}$ equal to $1$ and the other coefficients $\beta_i$ to $0$.

\begin{Rem}\label{RemSynchCaseAsLiuHughes}
The last three exponents are the same as the SMAC exponents $E_{rX}$, $E_{rY}$, $E_{rXY}$ of Liu and Hughes \cite{Hughes}, in absence of a time sharing variable $U$.  For AMAC, time sharing is possible only in case of known delay. Then it could improve error exponents like for SMAC, but this is out of the scope of the paper.
\end{Rem}

Recall that $d$ and $l$ are defined by (\ref{Defd}), (\ref{DeflSzam}), and $\delta_n$ by (\ref{EqDeltan}). 
\begin{Thm}[Main result]\label{ThmRovid}
  For all $K$, $n$, types $P^X\in \iP^n(\iX)$, $P^Y\in \iP^n(\iY)$, and rates $R_1\leq \HH(P^X)-\delta_n$, $R_2\leq \HH(P^Y)-\delta_n$ there exists an AMAC code as in Definition \ref{DefCode} such that for each channel matrix $W$ and delay $D$ the probability of each error pattern $\vS$ in the first passage of this section is bounded as
  \[P_e^D(\vS)\leq 2^{-n(E^\alpha(L,j)-\gamma_n)},\quad \alpha=\frac{d}{n}\]
  and consequently
  \begin{equation} \label{EqFinalestForm}
    P_e^D\leq 2^{-n(E^{\alpha,M}-\gamma_n)}, \quad \alpha=\frac{d}{n}, \quad M= 2 \max (l-1,K-l),
  \end{equation} 
  where $\gamma_n\to 0$ depends only on $|\iX|$, $|\iY|$, $|\iZ|$ and $K$. Each $E^\alpha (L,j)$ is a (uniformly) continous function of $(\alpha, P^X,P^Y,R_1,R_2,W)$ and so is also $E^{\alpha,M}$. Moroever, each $E^\alpha (L,j)$ is a jointly convex function of $(R_1,R_2,W)$ when $\alpha$, $P^X$, $P^Y$ are fixed.  
\end{Thm}
\begin{IEEEproof}
The existence of AMAC codes as claimed follows from Theorems \ref{ThmExpEgyenletesFolyt} , \ref{ThmConvexityFinalform} using (\ref{eqLessl}). Formally, one has to check that the needed results hold also in the previously excluded case $d=0$, with suitable minor modifications of the deifnitions and proofs. This is left to the reader. 

The continuous dependence of the exponents on $(\alpha, P^X,P^Y, R_1,R_2,W)$ will be proved in Appendix B. 

Finally, rewriting the $|\cdot|^+$ term in (\ref{ELJ}) as in the proof of Theorem \ref{ThmConvexityFinalform}, one sees that for fixed $\alpha$, $P^X$, $P^Y$ the expression to be minimized is jointly convex in $V_1$, $V_2$, $V_{12}$ (subject to the constraints $V_1$, $V_2$, $V_{12} \in \iP^*$) and $R_1$, $R_2$, $W$. This implies that the minimum $E^{\alpha} (L,j)$ is a jointly convex function of $R_1$, $R_2$, $W$ as claimed.  \end{IEEEproof}

\begin{Rem} \label{remarkdegeneralteset}
 In the degenerate cases $R_1=0$, $R_2=0$ a simple improvement of the bound \eqref{EqFinalestForm} may be possible. If $R_2=0$, say, then no errors can occur for sender 2 thus the possible irreducible error patterns are those that represent a single error for sender 1. This implies that $E^{\alpha,M}$ in \eqref{EqFinalestForm} may be replaced by $E^\alpha(1,1)$, even if $(L,j)=(1,1)$ does not attain the minimum in \eqref{DefAsExp}.
\end{Rem}

In the next corollary, $R_1$, $R_2$ and $K$ are considered fixed, and $P^X$, $P^Y$ are chosen by the senders, depending or not on the channel $W$ according as they know it or not. 

When the delay $D$ is unknown to the senders, the performance criterion for an AMAC code is the worst case error probability
\begin{equation}
P_e^{worst}\triangleq \max_{0 \le D \le nK-1} P_e^D.
\end{equation}
When $D$ is chosen by the senders, the performance criterion is 
\begin{equation}
P_e^{best}\triangleq \min_{0 \le D \le nK-1} P_e^D,
\end{equation}
provided that the senders know $W$ (since the minimizer $D$ may depend on $W$, even if $P^X$, $P^Y$ are fixed).
\begin{Cor}\label{CorAsCasErrExp}
For any $P^X \in \iP(\iX)$, $P^Y \in \iP(\iY)$, there exist AMAC codes that guarantee for all channels $W$	
  \begin{align}
    &P_e^{worst}\leq \exp \left\{-n\left(\min\limits_{\alpha\in[0,1]}E^{\alpha, 2K-2}-\gamma_n\right)\right\} \label{EqWorstBound} \\
    &P_e^{best}\leq \exp \left\{-n\left(\max\limits_{\alpha\in[0,1]}E^{\alpha,K}-\gamma_n\right)\right\}. \label{EqBestBound}
  \end{align}
 When the senders and receiver know $W$, the choice of $P^X$, $P^Y$ may be tailored to $W$, achieving 
   \begin{align}
 &P_e^{worst} \leq \exp \left\{-n\left(\max_{P^X\in \iP(\iX),P^Y\in\iP(\iY)}\min\limits_{\alpha\in[0,1]}E^{\alpha,2K-2}-\gamma_n\right)\right\} \label{EqWorstBoundWithMaxInput} \\
 &    P_e^{best}\leq \exp \left\{-n\left(\max_{P^X\in \iP(\iX),P^Y\in\iP(\iY)}\max\limits_{\alpha\in[0,1]}E^{\alpha,K}-\gamma_n\right)\right\}. \label{EqBestBoundWithMaxInput}
 \end{align}
\end{Cor}
\begin{IEEEproof}
A direct application of Theorem \ref{ThmRovid} gives these bounds with maxima restricted to types $P^X \in \iP^n(\iX)$, $P^Y \in \iP^n(\iY)$ and $\alpha$ of form $\frac{d}{n}$. Note that in (\ref{ELJ}) any $l \in [K]$ may occur when $D$ is unknown, while if $D$ is chosen by the senders then $l=\lceil \frac{K}{2} \rceil$ may be taken to minimize $M = 2 \max (l-1,K-l)$. By continuity the maxima may be taken for all $P^X \in \iP(\iX)$, $P^Y \in \iP(\iY)$ and $\alpha \in [0,1]$, at the expense of admitting a larger $\gamma_n$ that still has the properties in Theorem \ref{ThmRovid}.
\end{IEEEproof}

\begin{Rem} \label{Remkimittud}
	While the bound (\ref{EqWorstBound}) is achievable even if senders and receiver are ignorant about the channel matrix $W$, the senders do need side information about $W$ to achieve (\ref{EqBestBound}), (\ref{EqWorstBoundWithMaxInput}), (\ref{EqBestBoundWithMaxInput}) via tayloring to it their choice of $P^X$, $P^Y$ (and of $D$, when chosen by them). The amount of this side information need not be more than constant times $\log n$ bits, since the number of possible choices is polynomial in $n$. In order to achieve (\ref{EqWorstBoundWithMaxInput}), (\ref{EqBestBoundWithMaxInput}) via the approach in this paper, the receiver also needs side information about $W$. Indeed, although the decoder in Definition \ref{DefDekodolo}  does not directly depend on $W$, it does depend on the employed codebooks that do depend on $W$ through the senders' choice of the codeword types $P^X$, $P^Y$. A more refined approach may eliminate the need for any side information about $W$ at the receiver. This is suggested by the known fact that the random coding exponent for single user channels is achievable even if the sender may use any codebook, unknown to the receiver, from a known collection of subexponentially many codebooks, see \cite{CsiszarJSCEE} and a generalization to SMAC in \cite{RAC}.
\end{Rem}

The next theorem characterizes the positivity of the exponents $E^{\alpha} (L,j)$ and facilitates their numerical evaluation. It is an analogue of a standard result about the random coding exponent function of single-user channels, see Corollary 10.4 and eq. (10.23) of \cite{Csiszar}. The role there of $R$ and $\hat{R}$ is played by the linear combinations
\begin{equation}
r (R_1,R_2) \triangleq (\beta_1 + \beta_{12})  R_1 + (\beta_2 + \beta_{12}) R_2 = \sum_{i \in \{1,2,12\}} \beta_i R_i,  \label{equation-l}
\end{equation}
with $\beta_1$, $\beta_2$, $\beta_{12}$ defined as in (\ref{ELJ}), and by $\hat{r}$ defined below.  

Given $P^X$, $P^Y$ and $W$, write $P=P^{XYZ}$, and denote by $V_i^*$ the (unique) minimizer of $\DD (V||P^{XYZ}) + \I_{V}^i$ subject to $V \in \iP^*, i \in \{1,2,12\}$. Further, denote 
\begin{equation}
\hat{r} \triangleq  \sum_{i \in \{1,2,12\}} \beta_i \I_{V_i^*}^i. 
\end{equation}
Note that the distirbutions $V_i^*$ do not depend on $\alpha$, $L$, $j$, while $r (R_1,R_2)$ and $\hat{r}$ do through the parameters $\beta_i$.    
\begin{Thm}\label{ThmOptimalDist}
The exponent $E^\alpha (L,j)$ depends on $R_1$, $R_2$ through $r (R_1,R_2)$, and is a convex function of $r (R_1,R_2)$. It is positive if and only if 
\begin{equation}
r (R_1,R_2)  < \sum_{i \in \{1,2,12\}} \beta_i \I_{P}^i \label{equation-po}
\end{equation}
and then equals
\begin{enumerate}[label=(\roman*)]
	\item  $\sum_{i \in \{1,2,12\}} \beta_i  \left[  \DD (V_i^*||P^{XYZ}) + \I_{V_i^*} \right] - r (R_1,R_2)$, if $r (R_1,R_2) \le \hat{r}$
	\item the minimum of $\sum_{i \in \{1,2,12\}} \beta_i  \DD (V_i||P^{XYZ})$ for $V_1$, $V_2$, $V_{12}$ in $\iP^*$ satisfying $\sum_{i \in \{1,2,12\}} \beta_i \I_{V_i}^i = r (R_1,R_2)$, if $r (R_1,R_2) > \hat{r}$.
\end{enumerate}	
 \end{Thm}

  
  

\begin{IEEEproof} 
Since
\begin{equation}
\sum_{i \in \{1,2,12\}} \beta_i  ( \I_{V_i}^i - R_i) = \sum_{i \in \{1,2,12\}} \beta_i  \I_{V_i}^i -r (R_1,R_2),  \label{equation-bl}
\end{equation}
the first assertion (convexity) is obvious from (\ref{ELJ}) and the convexity assertion of Theorem \ref{ThmRovid}. 

The expression minimized in (\ref{ELJ}) is positive unless $V_1=P$ for each $i \in \{1,2,12\}$ with $\beta_i >0$ and, in addition, the quantity in (\ref{equation-bl}) is nonpositive. This proves the second assertion (positivity). The claim (i) immediately follows from (\ref{ELJ}), definition of $\hat r$ and (\ref{equation-bl}). 

The claim (ii) will be established if we show that in case  $\hat{r} < r (R_1,R_2) < \sum_{i \in \{1,2,12\}} \beta_i \I_{P}^i$ the quantity in (\ref{equation-bl}) is equal to $0$ for $(V_1,V_2,V_{12})$ attaining the minimum in (\ref{ELJ}). If it were negative then its positive part (equal to $0$) would not change by a sufficiently small increase of $R_1$ and $R_2$, yielding a contradiction to the consequence of the first two assertions that $E^\alpha (L,j)$ as a function of $r (R_1,R_2)$ strictly decreases subject to \eqref{equation-po}\footnote{Alternatively one could prove this assertion ---that \eqref{equation-bl} cannot be negative--- by taking a convex combination with $P^{XYZ}$ and use a similar train of thoughts as the next one.}. 

If the expression in (\ref{equation-bl}) were positive then for $s$ sufficiently small also
\begin{equation}
 \sum_{i \in \{1,2,12\}} \beta_i  \I_{(1-s) V_{i} + s V_{i}^*}^i  -r (R_1,R_2)
\end{equation}
would be also positive. Then the following chain of inequalities implied by convexity and the definition of $(V_1^*, V_2^*, V_{12}^*)$ would contradict the fact that $(V_1,V_2,V_{12})$ attains the minimum in (\ref{ELJ}):
\begin{align}
 &\sum_{i \in \{1,2,12\}} \beta_i \left[ \DD\left((1-s) V_{i} + s V_{i}^*\|P^{XYZ}\right) + \I_{(1-s) V_{i} + s V_{i}^*}^i  \right] \notag\\
&< (1-s) \sum_{i \in \{1,2,12\}} \beta_i \left[ \DD( V_{i} \|P^{XYZ}) + \I_{V_{i}}^i  \right] +  s \sum_{i \in \{1,2,12\}} \beta_i \left[ \DD(  V_{i}^*\|P^{XYZ}) + \I_{V_{i}^*}^i  \right]  \label{nullasagbizpozeset1} \\
&< \sum_{i \in \{1,2,12\}} \beta_i \left[ \DD( V_{i} \|P^{XYZ}) + \I_{V_{i}}^i  \right].  \label{nullasagbizpozeset2}
\end{align}
Here the inequalities are strict since $\hat{r} < r (R_1,R_2)$ excludes that $V_i=V^*_i$ for each $i \in \{1,2,12\}$ with $\beta_i >0$. \end{IEEEproof}

\begin{Rem} \label{Remarkelesegesseg}
Theorem \ref{ThmOptimalDist} implies that a sufficient condition for the positivity of each $E^{\alpha} (L,j)$ and hence also of the exponent  $\min_{\alpha \in [0,1]} E^{\alpha,2K-2}$ in (\ref{EqWorstBound}) is 
\begin{equation}
R_1 < \I_P^1, \quad R_2 < \I_P^2, \quad R_1 + R_2 < \I_{P}^{12}. \label{equation-int}
\end{equation}
When $K$ is large, this is close to the necessary condition that the effective rate pair has to be in $\iR(P^X,P^Y,W)$ defined by $(\ref{DefPentagon})$, i.e., 
	\begin{equation}
	\left( R_1\left(1-\frac{1}{K}\right),R_2\left(1-\frac{1}{K}\right)\right) \in \iR(P^X,P^Y,W) \label{equation-inR}.
	\end{equation}
	The necessity of (\ref{equation-inR}) follows from the fact that for constant composition block codes of increasing blocklength, without synch sequences, the worst case error probability can not approach $0$ if the codeword types approach $P^X$, $P^Y$ and the rate pairs approach some $(\hat{R}_1, \hat{R}_2) \notin \iR(P^X,P^Y,W)$ (see the proof of the converse part of AMAC coding theorem in \cite{Poltyrev}, \cite{AMAC}). Indeed, to each AMAC code as in Definition \ref{DefCode} there corresponds a block code of blocklength $nK$ and rate pair  $\left( R_1\left(1-\frac{1}{K}\right),R_2\left(1-\frac{1}{K}\right)\right)$, whose codewords are concatenations of $K-1$ codewords of the former and a synch sequence. Thus, if $\min_{\alpha \in [0,1]} E^{\alpha,2K-2}$ were positive for some $(R_1,R_2)$ not satisfying (\ref{equation-inR}), Corollary \ref{CorAsCasErrExp} would contradict the above fact. 
\end{Rem}

The next corollary concerns the capacity region of compound AMACs, i.e., AMACs whose channel matrix is an unknown member of a known (perhaps infinite) family $\irott W$ of matrices $W: \iX \times \iY \rightarrow \iZ$. The representation below of this capacity region appears in Shrader and Ephremides \cite{Comp-AMAC}, for a model somewhat different from ours. There,  it has been left open whether all $(\hat{R}_1, \hat{R}_2)$ in that region indeed belong to the capacity region also when (as here) the receiver is assumed ignorant of $W$. Formally, whether it is true that for every $\delta >0$ there exist block codes of increasing blocklength (without synch sequences) whose rate pairs are in the $\delta$-neighborhood of $(\hat{R}_1,\hat{R}_2)$ and whose error probability approaches $0$ uniformly for all possible delays and channels $W \in \irott W$.

\begin{Cor}\label{AMACEE-compound-AMAC-capacity}
The capacity region of a compound AMAC equals the union over all $P^{X} \in \iP(\iX)$, $P^{Y} \in \iP(\iY)$ of the sets 
\begin{align}
&\iR(P^X,P^Y,\irott W) \triangleq  \bigcap_{W\in \irott W}\iR(P^X,P^Y,W) = \left\{(\hat{R}_1,\hat{R}_2): 0 \le \hat{R}_1 \le \inf_{W \in \irott W} \I^1_{P^{XY} \circ W}, \right. \\
&\left. 0 \le \hat{R}_2 \le \inf_{W \in \irott W} \I^2_{P^{XY} \circ W}, \text{ }   \hat{R}_1 + \hat{R}_2 \le \inf_{W \in \irott W} \I^{12}_{P^{XY} \circ W} \right\}, \quad P^{XY}(x,y)=P^X (x) P^Y(y). \notag
\end{align}	
\end{Cor}
\begin{IEEEproof} 
We have to prove the claim preceding the Corollary for each $P^X$, $P^Y$ and $(\hat{R}_1,\hat{R}_2) \in \iR(P^X,P^Y,\irott W)$. Fixing $P^X$, $P^Y$ suppose first that $\hat{R}_1$ and $\hat{R}_2$ are both positive; then $\delta < \min (\hat{R}_1,\hat{R}_2)$ may be assumed. 

By Theorem \ref{ThmRovid}, $E^{\alpha,2K-2}$ is a continous function of $\alpha, R_1,R_2,W$, hence, its minimum over the compact set of all $(\alpha, R_1,R_2,W)$ with $\alpha \in [0,1]$ and 
\begin{equation}
0 \le R_1 \le  \I^1_{P^{XY} \circ W} - \frac{\delta}{2}, \text{ }   0 \le R_2 \le  \I^2_{P^{XY} \circ W} - \frac{\delta}{2}, \text{ }   R_1 + R_2 \le  \I^{12}_{P^{XY} \circ W} - \delta \label{equationcs}
\end{equation}
is attained. This minimum is positive by Theorem \ref{ThmOptimalDist}. The condition (\ref{equationcs}) do hold for each $W \in \irott W$ if $R_1 \le \hat{R}_1 -\frac{\delta}{2}$, $R_2 \le \hat{R}_2 -\frac{\delta}{2}$. Then, by Corollary \ref{CorAsCasErrExp}, there exist AMAC codes with rate pair $(R_1,R_2)$ whose error probability approaches $0$ uniformly for all delays and all channels $W \in \irott W$. If $K$ is large enough, $(R_1,R_2)$ can be chosen such that the effective rates $R_1 \left(1-\frac{1}{K}\right)$, $R_2 \left(1-\frac{1}{K}\right)$ exceed $\hat{R}_1 - \delta$ resp. $\hat{R}_2 - \delta$, thus the rate pairs of the block codes corresponding to these AMAC codes as in Remark \ref{Remarkelesegesseg} are in the $\delta$-neighborhood of $(\hat{R}_1, \hat{R}_2)$. 

The case of $\hat{R}_1$ or $\hat{R}_2$ equal to $0$ need not be considered when $\iR(P^X,P^Y,\irott W)$ is non-degenerate, i.e., has nonempty interior. When it is degenerate, say $\inf_{W \in \irott W} \I^2_{P^{XY} \circ W}=0$ thus $\iR(P^X,P^Y,\irott W) = \{(\hat{R}_1,0): 0 \le \hat{R}_1 \le \inf_{W \in \irott W} \I^1_{P^{XY} \circ W}\} $, the above argument has to be modified, relying upon Remark \ref{remarkdegeneralteset}, as follows. When $P^X$, $P^Y$ and $R_2 =0$ are fixed, the minimum of $E^{\alpha} (1,1)$ (not actually depending on $\alpha$) over all $(R_1,W)$ with $0 \le R_1 \le \I^{1}_{P^{XY} \circ W} - \frac{\delta}{2}$ is attained and positive. To each $(\hat{R}_1, 0) \in \iR(P^X,P^Y,\irott W)$ and $0 < \delta < \hat{R}_1$ take $R_1$ with $R_1 < \hat{R}_1 - \frac{\delta}{2}$ and $R_1 \left(1-\frac{1}{K}\right) > \hat{R}_1 - \delta$. Then there exist AMAC codes with rate pair $(R_1,0)$ whose error probability approaches $0$ uniformly over $W \in \irott W$. The corresponding block codes will have the desired properties. \end{IEEEproof} 

This section is concluded by the result in the title that controlled asnychronism may outperform synchronism. 

\begin{Thm}\label{ThmCASBiggerThanSMAC}
For some MAC and certain rates of transmission, the error exponent $E^{\alpha,K}$ in Corollary \ref{CorAsCasErrExp}  achievable by a suitable choice of delay (giving rise to a suitable $\alpha=\frac{d}{n}$) exceeds the best possible error exponent for synchronous transmission with the same rates. 
\end{Thm}

This result will be proved in the next section, via numerical calculation for a specific MAC. We expect, however, that this phenomenon actually occurs for most MACs.

\section{Numerical Results}\label{SecNumRes}

\begin{figure}[hbt]
\begin{center}
\includegraphics[width=0.4\textwidth]{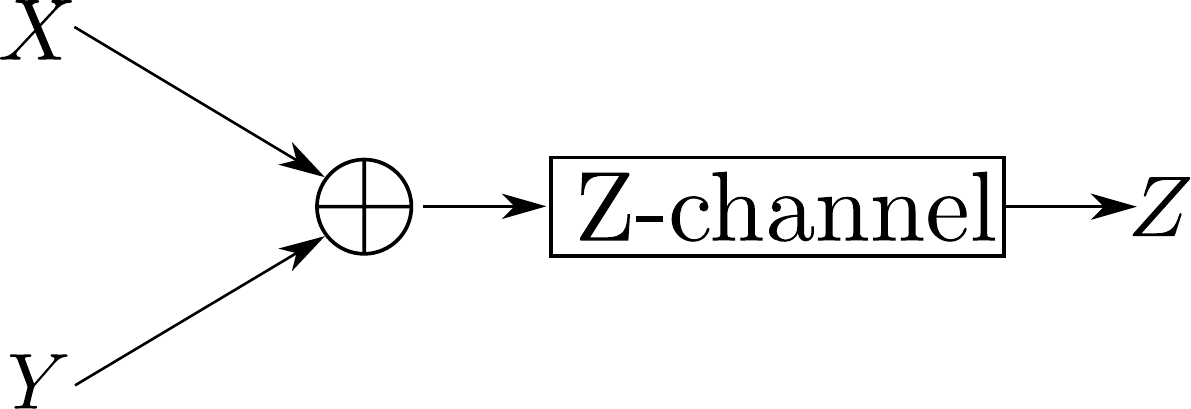}
\caption{ The MAC used for numerical calculations }\label{FigNumchannel}
\end{center}
\end{figure}

In this section, numerical results are reported, for the specific MAC depicted in Figure \ref{FigNumchannel}. It has binary alphabets $\iX=\iY=\iZ=\{0,1\}$ and the output is obtained by sending the $\mod$ 2 sum of the two inputs over a Z-channel. 

Formally, $W(z|x,y)=W_1(z|x\oplus y)$, $x,y,z\in\{0,1\}$ where 
\[W_1=\left(\begin{array}{rc} 1 & 0\\ \sigma & 1-\sigma \end{array} \right)\] is a Z-channel with cross over probability $\sigma$. In this section, all AMAC and SMAC error exponents are meant for this MAC.
 
 Synchronous transmission over the MAC in Figure \ref{FigNumchannel} has been treated in \cite{NazariUpper}. There, SMAC error exponents in \cite{Nazari} were shown tight in a sense, though apparently falling short of determining the reliability function at least for some rates. Motivated by \cite{Arunjavaslat}, here we use only the fact that SMAC block codes of rate pair $(R_1,R_2)$ give rise to single user block codes of rate $R_1+R_2$ for the Z-channel, more exactly its consequence that the SMAC reliability function is bounded above by $E_{sp}(R_1+R_2)$. Here $E_{sp}(R)$ denotes the sphere packing exponent of the Z-channel, see e.g. \cite[,Theorem 10.6]{Csiszar}, which is easy to evaluate numerically. Comparing the AMAC error exponent in Theorem \ref{ThmRovid} with this upper bound to the best possible synchronous exponent will prove Theorem \ref{ThmCASBiggerThanSMAC} 
 
 We have evaluated the AMAC exponent $E^{\alpha,M}$ for $\alpha=\nicefrac{1}{2}$ and $M=K$, corresponding to $l=\frac{K}{2}$, i.e., for delay $D=\frac{K-1}{2}n$, for equal codeword types $P^X=P^Y=P^*$ and equal rates $R_1=R_2=R$. In this case, the exponent $E^\alpha(L,j)$, $j\in\{1,2\}$ for irreducible error patterns of length $L$ do not depend on $j\in \{1,2\}$ due to symmetry, and are equal to
 \begin{align}
   E^{\nicefrac{1}{2}}(L) \triangleq \min_{V_1,V_{12}:V_1^X=V_1^Y=V_{12}^X=V_{12}^Y=P^*} & \left[ \DD(V_1\|P^{XYZ})+\frac{L-1}{2}\DD(V_{12}\|P^{XYZ})+\left|\I_{V_1}^1+\frac{L-1}{2}\I_{V_{12}}^2-LR\right|^+ \right]\label{EqEgyszeruNumExp}
 \end{align}
 where $P^{XYZ}(x,y,z)=P^*(x)P^*(y)W_1(z|x\oplus y)$. The latter is obvious when $L$ is odd. To verify it when $L$ is even, write $\DD(V_2\|P^{XYZ})$ and $\I^2_{V_2}$ in \eqref{ELJ} equivalently as $\DD(\hat V_2\|P^{XYZ})$ and $\I^1_{\hat V_2}$, where $\hat V_2 (z|x,y)\triangleq V_2 (z|y,x)$, and  use convexity as in the proof of Theorem \ref{ThmConvexityFinalform}. The actual computation has not used directly \eqref{EqEgyszeruNumExp} but the correspondig version of Theorem \ref{ThmOptimalDist}, with coefficients $\beta_1=1$, $\beta_2=0$, $\beta_{12}=\frac{L-1}{2}$, and with $r(R_1,R_2)=r(R,R)=LR$. 
 
 The exponents $E^{\nicefrac{1}{2}}(L)$, $L\in[K]$ and their minimum $E^{\nicefrac{1}{2},K}$ have been evaluated for $K=40$, in case of Z-channel crossover probability $\sigma=0.101$. This $Z$-channel had capacity $C(W_1)=0.761167$ attained for input distribution $Q=\{0.543959,1-0.543959\}$. The common input distribution $P^*$ for the two senders of the MAC has been chosen to make the mod 2 sum of two independent $P^*$ distributed random variables have distribution $Q$. This choice, numerically $P^*=\{0.351746,1-0.351746\}$, gave 
 \begin{align}
    \I^{12}_P=\I_P (XY\wedge Z)=\I_P(X\oplus Y,Z)=C(W_1). \label{EqNumIP}
 \end{align}
 
\begin{figure}[h!bt]
\begin{center}
\includegraphics[width=0.75\textwidth]{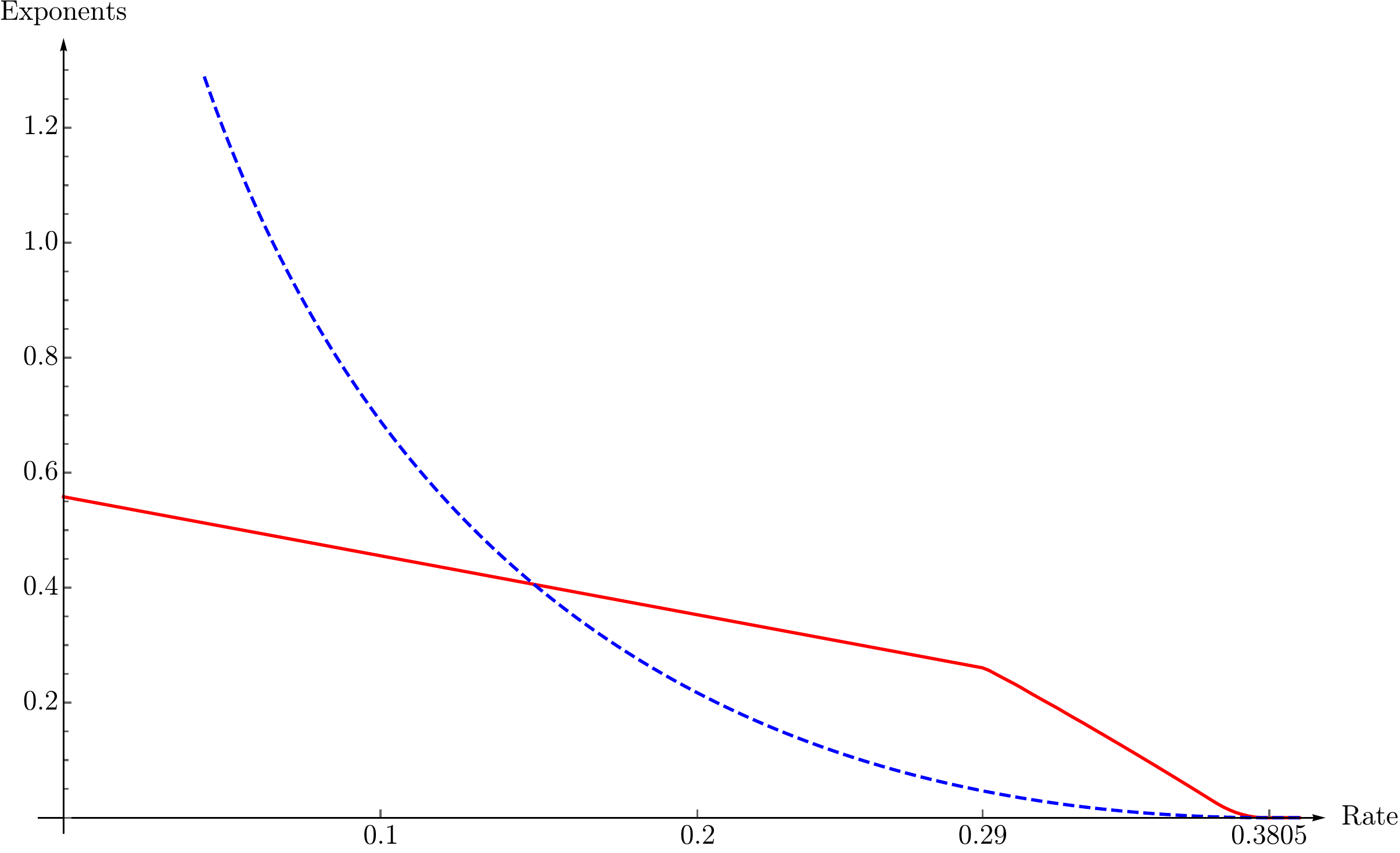} 
\end{center}
\caption{\small Controlled asynchronous exponent $E^{\nicefrac{1}{2},40}$ (continuous line) and upper bound to the best possible synchronous exponent (dashed line) as a function of the transmission rate, in the treated particular case. \label{FigExponents}}
\end{figure}

 The calculation has been carried out by Mathematica 11.2. The resulting error exponents $E^{\nicefrac{1}{2},40}=\min_{L \in [40]}E^{\nicefrac{1}{2}}$ are depicted in Figure \ref{FigExponents}, as a function the common rate $R$ of the senders, together with the upper bound $E_{sp}(2R)$ to the SMAC reliability function. For fair comparison, the AMAC exponent curve is drawn adjusting nominal rates to effective rates, i.e. as a function of the effective rate equal to $\frac{39}{40}$ times the nominal rate. 

 These numerical results prove Theorem \ref{ThmCASBiggerThanSMAC}.
 The supremum $R_{sup}$ of the rates $R$ for which $E^{\nicefrac{1}{2},40}>0$ is equal to $\approx0.37917$ in effective rate or $\approx0.38889$ in nominal rate. It differs only little from $\frac{1}{2}C(W_1)\approx 0.38058$, as it should according to Remark \ref{Remarkelesegesseg} that implies
 \[R_{sup}\left(1-\frac{1}{K}\right)=R^{eff}_{sup}<\frac{1}{2}C(W_1)<R_{sup}\]

\begin{figure}[!ht]
	\begin{center}
		\includegraphics[width=0.9\textwidth]{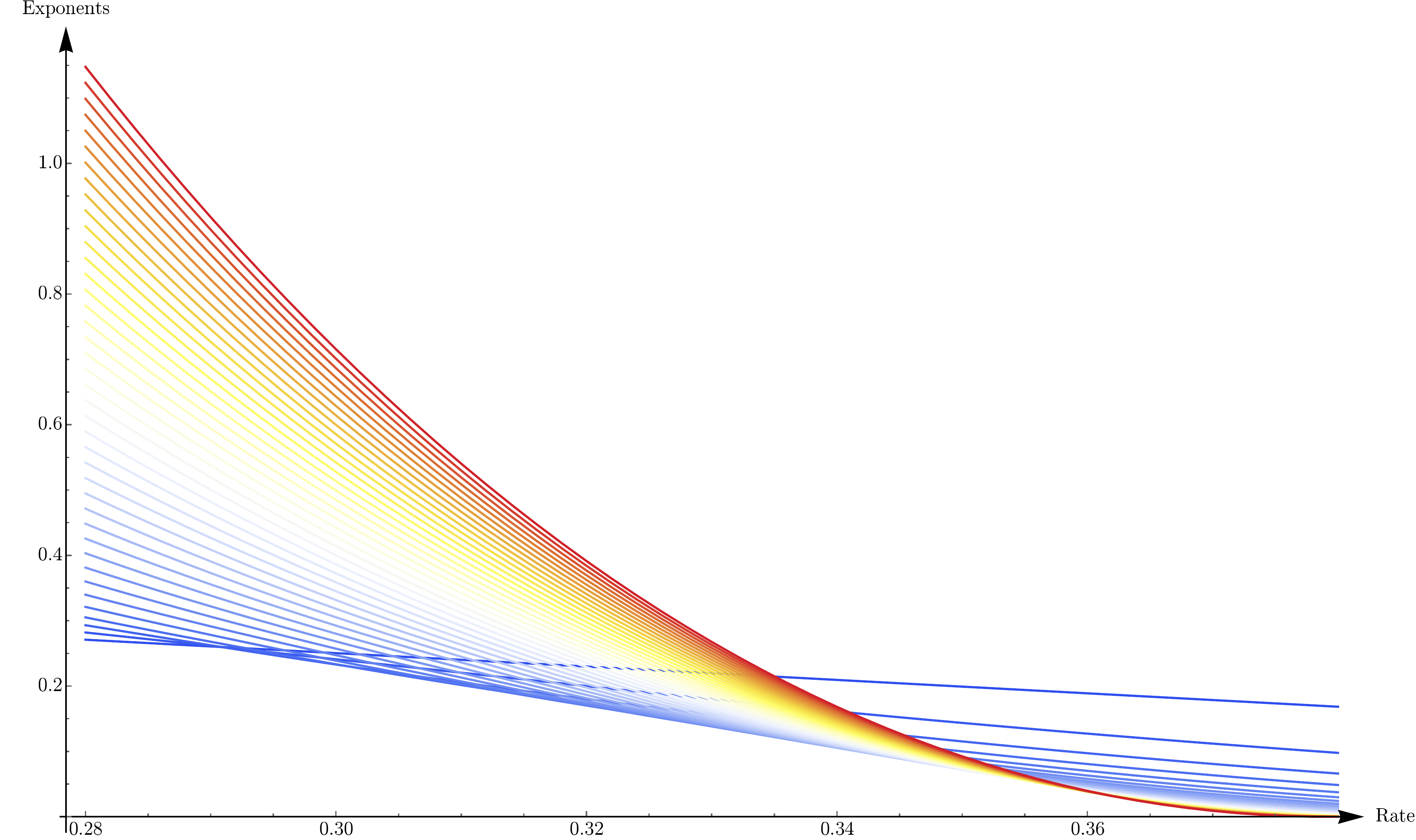} 
	\end{center}
	\caption{Exponents of error patterns of different lengths}\label{FigAllPattOnce}
\end{figure}

For $R\in [0,0.29]$ the exponent $E^{\nicefrac{1}{2},40}$ is equal to $E^{\nicefrac{1}{2}}(1)$, which is a linear function of $R$ in this interval, by Theorem \ref{ThmOptimalDist}. For larger $R$, $E^{\nicefrac{1}{2},40}$  is no longer linear, and is only piecewise convex  (though not visible for insufficient resolution) as the minimum of convex functions $E^{\nicefrac{1}{2}}(L)$, $L \in [40]$. The latter are depicted in Figure \ref{FigAllPattOnce}, for $R > 0.29$. 

\begin{figure}[hbt]
 \begin{center}
  \includegraphics[width=0.8\textwidth]{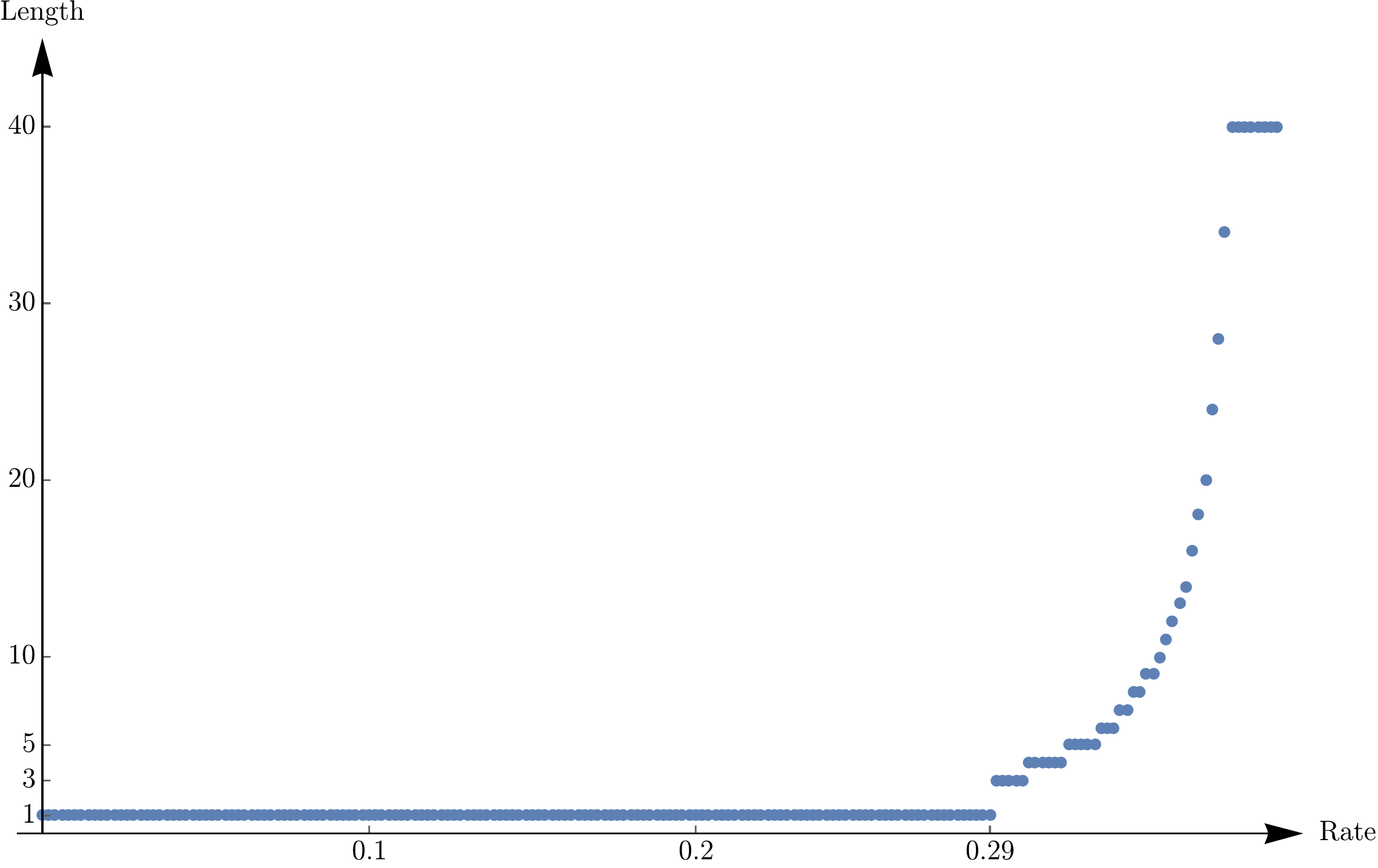} 
\end{center}
 \caption{Dominant error pattern lengths for different rates}\label{FigPattLength}
\end{figure}

 By dominant error pattern length $L_{dom}$ we mean $L \in [K]$ minimizing  $E^{\nicefrac{1}{2}}(L)$, i.e., attaining $E^{\nicefrac{1}{2}}(L_{dom})=E^{\nicefrac{1}{2},40}$. As intuition suggests, $L_{dom}$ increases with the rate $R$. It equals $1$ for $R\in [0,0.29]$ and takes the largest possible value $40$ for $R$ close to $R_{sup}$, see Figure \ref{FigPattLength}. The apparent jumps by more than 1 of $L_{dom}$ as a function of $R$ are likely caused by our having calculated $L_{dom}$ for rates growing by steps 0.002.

\section{Conclusions and outlook}\label{SecConclusion}

AMAC error exponents, i.e., error exponents for asynchronous transmission over multiple access channels have been derived, universally achievable in the limit of large blocklength, when the delay is unknown to the senders or chosen by them. The exponents have sufficiently simple  form for numerical computation. Our model involves a decoding window of lenght equal to $K$ times the codeword lenght $n$, as opposed to synchronous transmission via standard block codes when it would be pointless to use decoding window longer than $n$. Hence, AMAC error exponents could be meant in two different senses, relative to either the codeword length $n$ or the decoding delay $Kn$. In this paper, the first alternative has been adopted.

The method of subtypes has been employed, as in previous works of Farkas and Kói \cite{Hawaii}--\cite{Paris} with the exception of \cite{RAC}. A new technical tool has been the $\delta$-balanced sequences. One of the main technical result is the proof of packing lemma  (Lemma \ref{LemPacking-basic}). It overcomes a major obstacle to applying the method of random selection in cases when repetition of codewords may occur, namely that the codewords at different instances can not be independently chosen. This proof may open new perspectives also in other contexts, such as trellis code error exponents for single-user channels, as will be argued at the end of this section.

Another key result is the proof that controlled asynchronism, when senders transmit with a chosen delay, may be substantially more reliable than synchronous transmission. Though proven for a particular MAC, this is likely the rule rather than an exception. As a  heuristic reason, note that in the synchronous case the possible error patterns are that sender 1 or sender 2 or both are incorrectly decoded, and often the last one is the most likely. Then, transmission with chosen delay can be expected to decrease error probability, since it causes all error patterns to contain subblocks that are erroneous only for one sender. 

Many problems arising in the context of this paper could not be treated here. It would be most desirable to prove that our AMAC error bounds hold not only asymptotically but already for blocklengths in practice, and are achievable via a decoder whose computational complexity does not prohibit implementation. Apparently, this would require new methods. Other natural questions, however, can likely (or certainly) be settled by methods in this paper. We briefly survey some of them, to give an outlook. 

Improvements of the AMAC error exponents in Theorem \ref{ThmRovid} have not been addressed, but should be possible at least for small rates, via considerations familiar for single user channels. When the senders know the delay, the exponent in Theorem \ref{ThmRovid} could be improved via codeword selection from a conditional type class, conditioned on an auxiliary ``time sharing sequence'', as for SMAC in \cite{Hughes}, see Remark \ref{RemSynchCaseAsLiuHughes}. This could make sure for all rate pairs inside the capacity region of SMAC (rather than the perhaps smaller one of AMAC) that controlled asynchronous transmission admits positive error exponents, likely better than synchronous transmission. The error bounds \eqref{EqWorstBoundWithMaxInput}, \eqref{EqBestBoundWithMaxInput} in Corollary \ref{CorAsCasErrExp} proved for the case when senders and receiver know the channel matrix $W$ are likely achievable also when only the senders know $W$, see Remark \ref{Remkimittud}.

While the MMI decoder used here appears most suitable to obtain universal error bounds, alternate decoders are also of interest. When the channel matrix $W$ is known, maximum likelihood decoding may yield better exponents. The proof of Theorem \ref{ThmHawaiiTovabbfejlesztve} is not hard to modify for alternate decoders that share the property of simultaneously decoding all messages in the decoding window. On the other hand, Farkas and Kói \cite{ContSuccMAC} have analyzed the performance of successive decoding for AMAC, using subtypes as here but differently in details. They derived error exponents positive inside the capacity region but smaller than those in this paper. 

The extension of our results to more than two senders is beyond the scope of this paper. We only mention that it looks preferable to modify Definition \ref{DefCode} letting the senders use different periods $K$, relative prime to each other. This makes sure that a decoding window not splitting codewords can be found in case of any possible delays, its length is $n$ times the product of the periods. Then our derivations appear to carry over to more than $2$ senders without substantial changes.

From a mathematical, though less from an engineering point of view, a possible objection to our definition of AMAC code is that it implicitly assumes memory at the encoders (to know when to insert synch sequences). With somewhat more work, however, effectively the same exponents could be shown achievable also in a model not employing synch sequences, at least if the receiver's ability to locate codeword boundaries is still assumed. One could likely dispense with that assumption too, via substantially more work. We intend to return to this issue elsewhere. 

Finally, we point out the relationship of AMAC codes to trellis codes for single-user channels. A good early reference to trellis code error exponents is Forney \cite{Forney}. 

Focusing for symplicity to the special case treated in section \ref{SecMainResult} to any AMAC code we can assign a trellis code as follows.  Suppose $n$ is even, consider messages of form $m_t=(i_t,j_t) \in [2^{nR}] \times [2^{nR}]$, and let $\vy'(j)$ and $\vy''(j)$ denote the first and second half of the AMAC codeword $\vy(j)$ (or synch sequence if $j=0$). Encode blocks of $K-1$ messages $m_1,\dots,m_{K-1}$ by concatenations of $K$ blocks $\vx(i_t)\oplus \vy''(j_{t-1})\vy'(j_{t}) \in \{0,1\}^n$, $t\in [K]$, where $i_0=j_0=i_K=j_K \triangleq 0$. With the terminology of \cite[Definition 4]{Forney}, this defines an $(M,v,n,\tau)$ terminated trellis code with $M=2^{2nR}$, $v=1$, $\tau=K$ (apart from the unsubstantial detail that $m_0=m_K=(0,0)$ is not included in the message set $[2^{nR}] \times [2^{nR}]$). This trellis code is time-invariant, whereas if multiple codebooks were admitted for AMAC, the corresponding trellis code would be time-varying. 

As observed in \cite{Forney}, the principal trellis code error exponent results have been proved for time varying trellis codes, there has been no success in proving them for time-invariant ones. Clearly, the reason has been that repetitions cause a technical obstacle in the time-invariant case. To our knowledge, this obstacle has not been overcome since then. Also the recent work Merhav \cite{MerhavTrellis}, addressing expected values of random trellis codes, considers ensembles of time-varying trellis codes. 

The approach in this paper appears suitable for trellis codes also beyond those that correspond to AMAC codes. We expect that ideas as in the proof of our packing lemma will admit to overcome the obstacle of repetitions, and dispense with time-varying trellis codes, as we were able to dispense with multiple codebooks for AMAC.

\begin{appendices}

 \section{Proof of the Packing lemma} \label{Appendixpacking}

The proof of Lemma \ref{LemPacking-basic} uses random selection, but message sequences with repetitions cause a substantial technical difficulty. To overcome this obstacle, additional artificial packing inequalities will be considered. They involve \emph{mutilated message sequences} $\vi=(i_1,\dots,i_K)$, $\vj=(j_1,\dots,j_K)$ obtained replacing some components of a message sequence as in Section II by the symbol $e$, interpreted as erasing that component. 

The \emph{support} of a mutilated message sequence $\vi$ or $\vj$ is the set of indices $t \in [K]$ with $i_t$ or $j_t$ not equal to $e$. For $\vL =(L_X, L_{\hat X},L_Y,L_{\hat Y})$ a quadruple $(\vi,\hat \vi,\vj,\hat \vj)$ will be called $\vL$-admissible, denoted by $(\vi,\hat \vi,\vj,\hat \vj) \in \mathcal{A}(\vL)$, if $\vi$, $\hat \vi$, $\vj$, $\hat \vj$ have supports $L_X$, $L_{\hat X}$, $L_Y$, $L_{\hat Y}$ and, in addition, $i_t \ne \hat i_t$ if $t \in L_X \cup L_{\hat X}$, $j_t \ne \hat j_t$ if $t \in L_Y \cup L_{\hat Y}$.

\begin{Rem} \label{remarkerrorauxpatt}
	This concept, though not intuitively motivated, includes error patterns as special cases in the following sense. If $(\vi,\hat \vi,\vj,\hat \vj) \in \iE\iP(L_1,L_2)$ then, erasing from  $\hat \vi$ resp. $\hat \vj$ the components  $\hat i_t$ resp. $\hat j_t$ with $t \notin L_1$ resp. $t \notin L_2$ (formally, replacing them by $e$), the resulting mutilated sequences $\hat \vi'$ and $\hat \vj'$ satisfy $(\vi,\hat \vi',\vj,\hat \vj') \in \mathcal{A}([K],L_1,[K],L_2)$. Conversely, each quadruple in $\mathcal{A}([K],L_1,[K],L_2)$ arises uniquely in this way.
\end{Rem}
As repetitions in message sequences cause a major technical problem, they need special attention. A quadruple $(\vi,\hat \vi,\vj,\hat \vj) \in \mathcal{A}(\vL)$ will be said to have \emph{repetition pattern} $\mathbf{\tilde{L}}=(\tilde{L}_X,\tilde{L}_{\hat{X}},\tilde{L}_Y,\tilde{L}_{\hat{Y}})$ if
\begin{align}
&\{t \in L_X: i_s=i_t \text{ or } \hat{i}_s=i_t \text{ for some } s<t \} = \tilde{L}_X	\\
& \{t \in L_{\hat{X}}: i_s=\hat{i}_t \text{ or } \hat{i}_s=\hat{i}_t \text{ for some } s<t \} = \tilde{L}_{\hat{X}},
\end{align}
and similarly for $\tilde{L}_Y$, $\tilde{L}_{\hat{Y}}$. The set of possible repetitions patterns of $\vL$-admissible quadruples is denoted by $Rep(\vL)$, and for $\vL$-admissible quadruples with repetition pattern $\mathbf{\tilde{L}} \in Rep(\vL)$ we write $(\vi,\hat \vi,\vj,\hat \vj) \in \mathcal{AR}(\vL, \mathbf{\tilde{L}} )$.   

For mutilated message sequences $\vi$, $\vj$ we still define sequences $\vx(\vi)$ and $\vy(\vj)$ by \eqref{DefxMintIndexFv}, \eqref{DefyMintIndexFv}, setting $\vx(e)=\vy(e) \triangleq *^n$, where $*$ stands for empty space. Arranging such sequences corresponding to a quadruple $(\vi,\hat \vi,\vj,\hat \vj)$ in a four-row array, subblocks and subtypes are defined as in Section II, now with subtypes $V_k \in \iP^{n_k}(\iX^* \times \iX^* \times \iY^* \times \iY^*)$ where $\iX^* \triangleq \iX \cup \{*\}$, $\iY^* \triangleq \iY \cup \{*\}$. 

\begin{figure}[hbt]
	\begin{center}
		\includegraphics[width=0.55\textwidth]{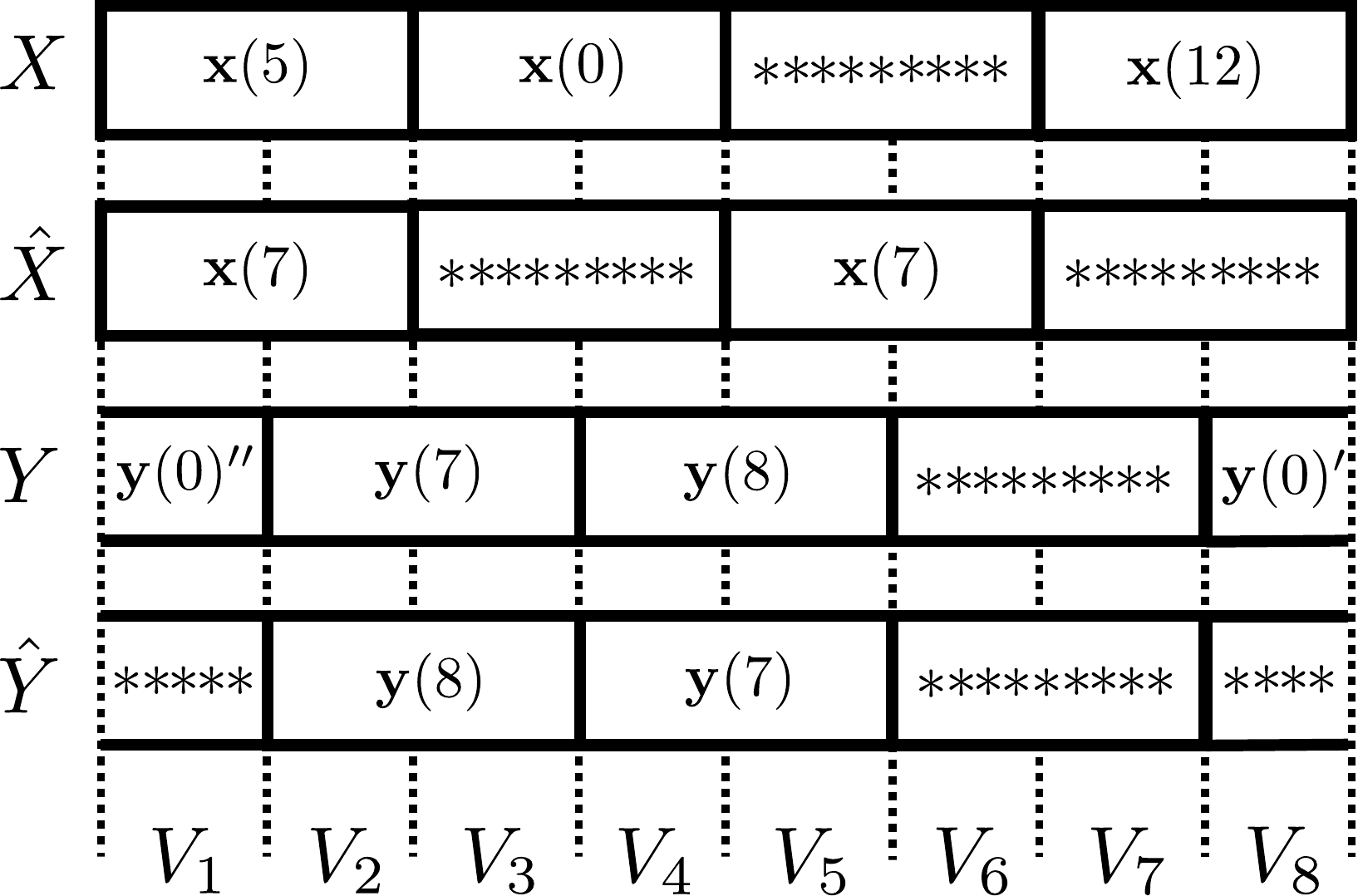}
	\end{center}
	\caption{Array corresponding to an $\vL$-admissible quadruple of mutilated message sequences for $\mathbf{L} = (\{1,2,4\},\{1,3\},\{1,2,3\},\{2,3\})$ with repetition pattern $\tilde{\mathbf{L}} = (\emptyset,\{3\},\{3\},\{3\})$. } \label{FigPacArr}
\end{figure}

For $\vL =(L_X, L_{\hat X},L_Y,L_{\hat Y})$ and $k \in [2K]$, let $B_{\vL}(k)$ denote the set of those rows of an array corresponding to $(\vi,\hat \vi',\vj,\hat \vj') \in \mathcal{A}(\vL)$ in which the $k$'th subblock is non-empty, i.e., does not equal $*^{n_k}$. Recalling that the rows of the array are referred to as rows $X$, $\hat X$, $Y$, $\hat Y$, the set  $B_{\vL}(k)$ is also regarded as a set of dummy random variables, $B_{\vL}(k) \subset \{X,\hat X, Y, \hat Y\}$. Further we write
\begin{align}
&R_{\vL}(k) \triangleq  \mathds{1}\left\{k \notin \{2l,2l+1\}  \right\} \cdot R_1 \cdot \left|B_{\vL} (k)\cap\{X,\hat{X}\} \right| \notag\\
&+\mathds{1}\left\{k \notin \{1,2K\}  \right\} \cdot R_2 \cdot \left|B_{\vL}(k)\cap\{Y,\hat{Y}\} \right|,   \label{blockonkentirata}
\end{align}  
where the indicator functions are necessary due to the distinguished role of the synchronization blocks, see (\ref{DeflSzam}) and the paragraph following it. For example in Fig. \ref{FigPacArr}, $B_{\vL}(4)=\{X, Y, \hat{Y}\}$ and $R_{\vL}(4)=2 R_2$.  
Note that
\begin{equation} \label{esetszam}
|\mathcal{A}(\vL)| = \exp \left\{\sum_{k \in [2K]} n_k R_{\vL}(k)\right\}
\end{equation}
and
\begin{equation} \label{esetszam2}
|\mathcal{AR}(\vL, \mathbf{\tilde{L}} )| < c(K) \cdot \exp \left\{\sum_{k \in [2K]} n_k (R_{\vL}(k) - R_{\mathbf{\tilde{L}}}(k))\right\};
\end{equation}
a suitable (crude) choice of the constant factor in \eqref{esetszam2} is $c(K)=(2K)^{4K}$.

For subsets $A$, $B$ of $\{X,\hat{X}, Y, \hat{Y}\}$, let $\I_V(A)$, $\HH_V(A)$ and $\HH_V(A|B)$ denote the multi-information, entropy and conditional entropy of the dummy random variables in the indicated sets, when $V^{X\hat{X}Y\hat{Y}}=V$ (equal to $0$ if $A = \emptyset$). The proof of Lemma \ref{LemPrePackingLemma} will use that in case $A \cap B = \emptyset$
\begin{align} 
&\HH_V(A|B) - \sum_{ \beta \in A} \HH_V (\beta) = \HH_V(A \cup B) - \HH_V(B) - \sum_{ \beta \in A\cup B} \HH_V (\beta)  + \sum_{ \beta \in B} \HH_V (\beta) =-\I_V(A \cup B) + \I_V(B).  \label{EqAlgebricmaniplemma2}
\end{align} 
For example, if $A= \{X,\hat{X}\}$ and $B=\{Y,\hat{Y}\}$ then \eqref{EqAlgebricmaniplemma2} says that
\begin{align} \label{EqSecificManip}
\HH_V(X,\hat{X}|Y,&\hat{Y}) - \HH_V ( X) - \HH_V (\hat{X}) =-\I_V(X \wedge \hat{X} \wedge Y \wedge \hat{Y})  + \I_V(Y \wedge \hat{Y}) .
\end{align} 

\begin{Lem}[Auxiliary packing lemma] \label{LemPrePackingLemma}
For each $K$, $n$, types $P^X\in\iP^n(\iX)$, $P^Y\in\iP^n(\iY)$, rates $R_1<\HH(P^X)-\delta_n,R_2<\HH(P^Y)-\delta_n$ and sets $\iT_1\subset \iT_{P^X}^n$, $\iT_2\subset \iT_{P^Y}^n$ of size not less than $ \frac{|\iT^n_{P^X}|}{2}$ resp. $\frac{|\iT^n_{P^Y}|}{2}$ there exists an AMAC code with codewords and synch sequences from $\iT_1$ resp. $\iT_2$ such that for each $D$,  $\vL =(L_X, L_{\hat X},L_Y,L_{\hat Y})$, $\tilde{\vL} \in Rep(\vL)$ and subtype sequence $\vV= (V_1, V_2,\dots, V_{2K})$ with $V_k \in \mathcal{P}^{n_i}(\mathcal{X}^* \times \mathcal{X}^* \times \mathcal{Y}^* \times \mathcal{Y}^*)$, $k \in [2K]$,  the following bound holds:
\begin{align} \label{statementpp}
&\sum_{(\vi,\hat\vi,\vj,\hat\vj) \in  \mathcal{AR}(\vL,\tilde{\vL}) )}\hspace{-4mm}\mathds{1}_{V_1, V_2,\dots, V_{2K}}\{\vx(\vi),\vx(\hat\vi),\vy(\vj),\vy(\hat\vj)\}\notag\\
& \leq p_n \cdot \exp\left\{ \sum_{k=1}^{2K} n_k \left[R_{\vL}(k)- R_{\mathbf{\tilde{L}}}(k)  -\left(\I_{V_k} (B_{\vL}(k))-\I_{V_k} (B_{     \mathbf{\tilde{L}}}(k))\right)\right]\right\}, 
\end{align}
where  $p_n$ is a polynomial of $n$ that depends only on $K$, $|\iX|$ and $|\iY|$. 
\end{Lem}
\begin{Rem}  \label{Remark0sag}
	In Lemma \ref{LemPrePackingLemma}, for each $(\vi,\hat \vi,\vj,\hat \vj) \in \mathcal{AR}(\mathbf{L},\tilde{\vL})$ the subtype sequence $\vV= (V_1, V_2,\dots, V_{2K})$ of the quadruple  $(\vx(\vi),\vx(\hat \vi),\vy(\vj), \vy(\hat \vj))$ is such that for each $k \in [2K]$ and $\beta \in \{X,\hat{X},Y,\hat{Y}\}$, the one dimensional marginal distributions $V_{k}^{\beta}$ are concentrated on $\mathcal{X}$ or $\mathcal{Y}$ if $\beta \in B_{\vL}(k)$ (according as $\beta \in \{X,\hat{X}\}$ or $\beta \in \{Y,\hat{Y}\}$), and on the symbol $*$ if $\beta \notin B_{\vL}(k)$. For each $t \in [K]$, if $\beta$ is in $\{X,\hat{X}\} \cap B_{\vL}(2t-1) = \{X,\hat{X}\} \cap B_{\vL}(2t)$ or in $\{Y,\hat{Y}\} \cap B_{\vL}(2t) = \{Y,\hat{Y}\} \cap B_{\vL}(2t+1)$ then the equality 
	\begin{equation} \label{auxiliarylemmabizhez}
	\frac{n_{2t-1}}{n}V_{2t-1}^{\beta} + \frac{n_{2t}}{n} V_{2t}^{\beta}=P^X \text{ or } \frac{n_{2t}}{n}V_{2t}^{\beta} + \frac{n_{2t+1}}{n} V_{2t+1}^{\beta}=P^Y
	\end{equation} 
	holds (when $2t+1=2K+1$, it is interpreted as $1$).    
\end{Rem}	
\begin{IEEEproof}  
	It is enough to prove the statement for subtype sequences with the properties in Remark \ref{Remark0sag}. Standard random coding argument is used with special attention to repetitions in the mutilated message sequences. Choose the codewords and synch sequences uniformly, without replacement, from  $\iT_1$ resp. $\iT_2$. For given $D$ (determining $d$ and $l$ by (\ref{Defd}), \eqref{DeflSzam})  let $\vX(\vi),\vX(\hat\vi),\vY(\vj),\vY(\hat\vj)$ denote the random sequences corresponding to a quadruple of mutilated message sequences  $(\vi,\hat \vi,\vj,\hat \vj) \in \mathcal{AR}(\vL, \mathbf{\tilde{L}} )$. The  number of possible realizations of $\vX(\vi),\vX(\hat\vi),\vY(\vj),\vY(\hat\vj)$ that has subtype sequence $\vV$ can be upper bounded by  
\begin{align}
&\prod_{k=1}^{2K} \exp \left\{ n_k \HH_{V_k}(B_{\vL}(k) \setminus B_{\mathbf{\tilde{L}}}(k)|B_{\mathbf{\tilde{L}}}(k))  \right\}. \label{elemszam}
\end{align}
Indeed, assume that the symbols corresponding to the first $k-1$ subblocks of $\vX(\vi),\vX(\hat\vi),\vY(\vj),\vY(\hat\vj)$ are fixed. Then the symbols of the $k$'th subblock of each row in $B_{\mathbf{\tilde{L}}}(k)$ are also determined. Hence, due to \eqref{EqBasicfact4}, the $k$'th factor in (\ref{elemszam}) upper-bounds the number of possible realizations of the symbols in the $k$'th subblock that yield subtype $V_k$, when the first $k-1$ subblocks are fixed.

As $(\vi,\hat \vi,\vj,\hat \vj) \in \mathcal{AR}(\vL, \mathbf{\tilde{L}} )$ implies that the pair of sequence $(\vi,\hat \vi)$ contains $|L_X\setminus \tilde{L}_X| + |L_{\hat{X}}\setminus \tilde{L}_{\hat{X}}|$ distinct indices in $[2^{nR_1}]$ and similarly for $(\vj,\hat \vj)$, each possible realization of $\left( \vX(\vi),\vX(\hat\vi),\vY(\vj),\vY(\hat\vj) \right)$ has probability
\begin{align} \label{valoszinusegkiirva}
&\left(|\iT_1| \cdot (|\iT_1|-1) \cdots (|\iT_1|- |L_X\setminus \tilde{L}_X| - |L_{\hat{X}}\setminus \tilde{L}_{\hat{X}}| +1) \right)^{-1} \notag\\
&\cdot  \left(|\iT_2| \cdot (|\iT_2|-1) \cdots (|\iT_2|- |L_Y\setminus \tilde{L}_Y| - |L_{\hat{Y}}\setminus \tilde{L}_{\hat{Y}}| +1) \right)^{-1}.
\end{align}

Here, each term of the first factor is bounded below by
\begin{equation}
\frac{1}{2} |\iT_1| \ge \frac{1}{4}|\iT_{P^X}^n| \ge \frac{1}{4} (n+1)^{-|\iX|} \cdot 2^{n\HH(P)},
\end{equation}
due to the consequence 
\begin{equation}
|L_X\setminus \tilde{L}_X| + |L_{\hat{X}}\setminus \tilde{L}_{\hat{X}}| \le 2^{nR_1} \le \frac{1}{4} |\iT_{P^X}^n| \le \frac{1}{2} |\iT_1|
\end{equation}
of $R_1<\HH(P^X)-\delta_n$. Of course, also $|L_X\setminus \tilde{L}_X| + |L_{\hat{X}}\setminus \tilde{L}_{\hat{X}}|\le 2K$. 

By these facts and their counterparts for $Y, \tilde{Y}$, (\ref{valoszinusegkiirva}) is bounded above by
\begin{align} \label{entropiasvalbecsles}
& 4^{2K} \cdot (n+1)^{ 2K(|\iX| +|\iY|)   } \notag\\
&\cdot \exp\left\{ -n(|L_X\setminus \tilde{L}_X| + |L_{\hat{X}}\setminus \tilde{L}_{\hat{X}}|) \HH(P^X) -n(|L_Y\setminus \tilde{L}_Y| + |L_{\hat{Y}}\setminus \tilde{L}_{\hat{Y}}|) \HH(P^Y)    \right\}.  
\end{align}

\eqref{auxiliarylemmabizhez} and the concavity of the entropy imply that \eqref{entropiasvalbecsles} can be further upper-bounded by  
\begin{align}
&4^{2K} (n+1)^{ 2K(|\iX| +|\iY|)   }  \exp \left\{- \sum_{k \in [2K]} n_k \sum_{\beta \in B_{\vL}(k) \setminus B_{\mathbf{\tilde{L}}}(k)} \HH_{V_k}(\beta)\right\}. 
\end{align}
Taking into account (\ref{EqAlgebricmaniplemma2}) it follows that the expectation of each term of the lhs of (\ref{statementpp}) (i.e., the probability of $\left( \vX(\vi),\vX(\hat\vi),\vY(\vj),\vY(\hat\vj) \right) \in \iT_{\vV}$) is bounded above by
\begin{align} \label{statementpp4}
4^{2K} (n+1)^{ 2K(|\iX| +|\iY|)   }  \exp \left\{ \sum_{k=1}^{2K} n_k \left[-\left(\I_{V_k} (B_{\vL}(k))-\I_{V_k} (B_{     \mathbf{\tilde{L}}}(k))\right)\right] \right\}.
\end{align}

Next, divide the lhs of (\ref{statementpp}) by its rhs without the factor $p_n$, and let $S$ denote the sum of the resulting expressions for all possible $D$, $\vL$, $\tilde{\vL}$ and $\vV$. The above considerations and (\ref{esetszam2}) imply that $\mathbb{E}(S) \le p_n$ for suitable $p_n (|\iX|,|\iY|,K)$. Hence, there exists a realization of the codebooks and the synch sequences with $S \le p_n$.
\end{IEEEproof}



\begin{Lem} \label{LemGenPackLemma}
	Any AMAC code as in Lemma \ref{LemPrePackingLemma} satisfies for each $D\in \{ 0,\dots,Kn-1 \}$, $\vL =(L_X, L_{\hat X},L_Y,L_{\hat Y})$ and subtype sequence $\vV=(V_1, V_2,\dots, V_{2K})$ with $V_k \in \mathcal{P}^{n_k}(\mathcal{X}^* \times \mathcal{X}^* \times \mathcal{Y}^* \times \mathcal{Y}^*)$, $k \in [2K]$ 
\begin{align} \label{statementgp}
&\sum_{(\vi,\hat\vi,\vj,\hat\vj) \in  \mathcal{A}(\vL)}\hspace{-4mm}\mathds{1}_{V_1, V_2,\dots, V_{2K}}\{\vx(\vi),\vx(\hat\vi),\vy(\vj),\vy(\hat\vj)\} \leq p''_n\cdot \exp \left\{ \sum_{k=1}^{2K} n_k \left[R_{\vL}(k)- \I_{V_k} (B_{\vL}(k))\right]\right\},
\end{align}
where  $p''_n = c(\vL) \left(p_n \right)^{|\vL|}$, $|\vL| \triangleq  |L_X| + |L_{\hat X}| + |L_Y| + |L_{\hat Y}|$. 
\end{Lem}
\begin{IEEEproof}
Fix a code and delay $D$ such that (\ref{statementpp}) holds for all $\vL$,  $\tilde{\vL} \in Rep(\vL)$ and  $\vV$. We prove the validity of (\ref{statementgp}) for all $\vL$ and $\vV$ by induction on $|\vL|$. It trivially holds if $|\vL|=0$. We claim that if (\ref{statementgp}) holds for all $\vV$ when $|\vL| \le m-1$ then it also holds when $|\vL| = m$.   
The lhs of (\ref{statementgp}) is equal to
\begin{align} \label{statementgp3}
	&\sum_{\tilde{\vL} \in Rep(\vL) } \hspace{-1mm} \sum_{\sumfrac{(\vi,\hat\vi,\vj,\hat\vj) }{\in \mathcal{AR}(\vL,\tilde{\vL})}}\hspace{-2mm}\mathds{1}_{V_1, V_2,\dots, V_{2K}}\{\vx(\vi),\vx(\hat\vi),\vy(\vj),\vy(\hat\vj)\}, \vspace{-4mm}
	\end{align}	
	
	Fix $\vL$ with $|\vL|=m$, $\tilde{\vL} \in Rep(\vL)$ and $\vV=(V_1, V_2,\dots, V_{2K})$. For each $k \in [2K]$ consider $\tilde{V}_k \in \mathcal{P}^{n_k}(\mathcal{X}^* \times \mathcal{X}^* \times \mathcal{Y}^* \times \mathcal{Y}^*)$ whose one-dimensional marginals $\tilde{V}_k^{\beta}$ with $\beta \notin B_{\tilde{\vL}}(k)$ are concentrated on $*$, and
	\begin{equation} \label{Vhullam}
\tilde{V}_k^{B_{\tilde{\vL}}(k)} = 	V_k^{B_{\tilde{\vL}}(k)}. 
\end{equation}
Then for each $(\vi,\hat\vi,\vj,\hat\vj) \in \mathcal{AR}(\vL,\tilde{\vL})$ that contributes to the inner sum in (\ref{statementgp3}), that is $(\vx(\vi),\vx(\hat\vi),\vy(\vj),\vy(\hat\vj)) \in \mathcal{T}_{\mathbf{V}}$, erasing the components of $\vi,\hat\vi,\vj,\hat\vj$ with indices not in $\tilde{L}_X$, $\tilde{L}_{\hat X}$, $\tilde{L}_Y$, $\tilde{L}_{\hat Y}$, respectively, changes the quadruple $(\vx(\vi),\vx(\hat\vi),\vy(\vj),\vy(\hat\vj))$ to one with subtype sequences $\mathbf{\tilde{V}} = (\tilde{V}_1,\tilde{V}_2,\dots,\tilde{V}_{2K})$, that is belonging to $\mathcal{T}_{\mathbf{\tilde{V}}}$. Since $|\tilde{\vL}| < |\vL|=m$, the induction hypothesis applied to $\tilde{\vL}$, $\mathbf{\tilde{V}}$ implies using (\ref{Vhullam}) that 
	\begin{align} 
	&\sum_{(\vi,\hat\vi,\vj,\hat\vj) \in  \mathcal{A}(\tilde{\vL})}\hspace{-4mm}\mathds{1}_{\tilde{V}_1, \tilde{V}_2,\dots, \tilde{V}_{2K}}\{\vx(\vi),\vx(\hat\vi),\vy(\vj),\vy(\hat\vj)\} \notag \\
		& \le c(\tilde{\vL}) \left(p(n) \right)^{|\tilde{\vL|}} \cdot \exp \left\{ \sum_{k=1}^{2K} n_k \left[R_{\tilde{\vL}}(k)- \I_{V_k} (B_{\tilde{\vL}}(k))\right]\right\}\label{statementgpind}
	\end{align}
	If the lhs of \eqref{statementgpind} is $0$ then also 
	\begin{align} \label{statementgp4}
	& \sum_{(\vi,\hat\vi,\vj,\hat\vj) \in  \mathcal{AR}(\vL,\tilde{\vL})}\hspace{-4mm}\mathds{1}_{V_1, V_2,\dots, V_{2K}}\{\vx(\vi),\vx(\hat\vi),\vy(\vj),\vy(\hat\vj)\} =0.
	\end{align}
	Otherwise, i.e., if the lhs of \eqref{statementgpind} is at least $1$, the inequality 
	\begin{align} \label{statementgpind2}
	& \exp \left\{ \sum_{k=1}^{2K} n_k \left[\I_{V_k} (B_{\tilde{\vL}}(k)) - R_{\tilde{\vL}}(k)\right]\right\} \le c(\tilde{\vL}) \left(p(n) \right)^{|\tilde{\vL|}}
	\end{align}
	holds. \eqref{statementgpind2} and the inequality \eqref{statementpp} in Lemma \ref{LemPrePackingLemma} imply that
	\begin{align} \label{statementgp5}
	& \sum_{(\vi,\hat\vi,\vj,\hat\vj) \in  \mathcal{AR}(\vL,\tilde{\vL})}\hspace{-4mm}\mathds{1}_{V_1, V_2,\dots, V_{2K}}\{\vx(\vi),\vx(\hat\vi),\vy(\vj),\vy(\hat\vj)\} \notag\\
	&\le p(n) \cdot c(\tilde{\vL}) \left(p(n) \right)^{|\tilde{\vL|}} \exp \left\{  \sum_{i=k}^{2K} n_k \left[R_{\vL}(k)- \I_{V_k} (B_{\vL}(k))\right]\right\} 
	\end{align}
	Then  \eqref{statementgp3}, \eqref{statementgp4} and  \eqref{statementgp5} prove \eqref{statementgp}. 	
\end{IEEEproof}
Lemma \ref{LemPrePackingLemma} and Lemma \ref{LemGenPackLemma} prove the existence of a code that satisfies (\ref{statementgp}) for each $D$, $\vL$ and $\vV$. Finally Lemma \ref{LemPacking-basic} follows from the instance $\vL= ([K],L_{1},[K],L_{2})$ of this result, using Remark \ref{remarkerrorauxpatt}.

\section{Continuity arguments} \label{Appendixfolyt}

In this Appendix, Theorem \ref{ThmExpEgyenletesFolyt} is proved and the proof of Theorem \ref{ThmRovid} is completed, via continuity arguments whose crucial ingredients are in Lemmas \ref{LemChangeDistribToVP} and \ref{LemContinuityTool} below. For better transparency, we will write briefly $V(x)$, $V(x,y)$, $V(z|x,y)$ etc. for $V^X(x)$, $V^{XY}(x,y)$, $V^{Z|XY}(z|x,y)$, etc. The conditional distribution $\{V(z|x,y), z \in \mathcal{Z}  \}$ is denoted by $V(\cdot|x,y)$. 

\begin{Lem}\label{LemChangeDistribToVP}
	To any distributions $V \in \iP (\iX \times \iY)$, $\hat{V}^X \in \iP (\iX)$, $\hat{V}^Y \in \iP (\iY)$ there exists $\hat{V} \in \iP (\iX \times \iY)$ whose marginals are the given $\hat{V}^X$, $\hat{V}^Y$, and
	\begin{equation}
	\| \hat{V} -V \| \le \|\hat{V}^X - V^X  \| + \|\hat{V}^Y - V^Y \| .
	\end{equation}
\end{Lem}
\begin{IEEEproof}
	We first contruct a distribution $\bar{V} \in \iP (\iX \times \iY)$ satisfying
	\begin{equation}
	\bar{V}^X = \hat{V}^X, \text{ } \bar{V}^Y = V^Y, \text{ }\| \bar{V} - V \| = \|\hat{V}^X - V^X \|.  \label{elsolepes}
	\end{equation}
	If $\hat{V}^X = V^X$ then $\bar{V}= V$ is suitable. Otherwise, denote the sets $\{x \in \iX: V (x) > \hat{V} (x)  \}$ and $\{x \in \iX: V (x) \le \hat{V} (x)  \}$ by $\iX^+$ and $\iX^-$, respectively, and define $\bar{V}$ by
	\begin{align}
	&\bar{V} (x,y) \triangleq 
	\left\{\begin{array}{lr}
	V(x,y)\frac{\hat{V}(x)}{V(x)} & \text{if }  x \in \iX^+\\[8pt]
	V(x,y) + c_y \cdot (\hat{V}(x)- V(x)) & \text{if }  x \in \iX^- \end{array} \right.,\\
	&c_y =\sum_{x' \in \iX^+} V(x',y) \frac{V(x')-\hat{V}(x')}{V(x')} \Big/ \sum_{x' \in \iX^-} (\hat{V}(x') - V(x')).     
	\end{align}
	The identity $\sum_{x' \in \iX^+} (V(x')-\hat{V}(x'))= \sum_{x' \in \iX^-} (\hat{V}(x')-V(x'))$ implies that $\sum_{y \in \iY} c_y =1$. Then (\ref{elsolepes}) follows by simple algebra. 
	
	Analogous construction with $\bar{V}$ in the role of $V$ and interchanging the roles of the two marginals gives a distribution  $\bar{\bar{V}} \in \iP (\iX \times \iY)$ satisfying
	\begin{equation}
	\bar{\bar{V}}^X = \bar{V}^X (= \hat{V}^X ), \text{ } \bar{\bar{V}}^Y = \hat{V}^Y, \text{ }\| \bar{\bar{V}} - \bar{V} \| = \|\hat{V}^Y - V^Y \|.  \label{masodiklepes}
	\end{equation}
	Due to (\ref{elsolepes}), (\ref{masodiklepes}) and triangle inequality, $\hat{V} \triangleq \bar{\bar{V}}$ satisfies the assertion of the lemma.  \end{IEEEproof}

\begin{Lem}\label{LemContinuityTool}
	Given any channel matrix $W$, and distributions  $V \in \iP (\iX \times \iY \times \iZ)$,  $\hat{V} \in \iP (\iX \times \iY)$ with $\| \hat{V}^{XY} -V^{XY} \| \le (|\iX||\iY|)^{-2}$, there exists $\hat{V} \in \iP (\iX \times \iY \times \iZ)$ whose $XY$-marginal is the given $\hat{V}^{XY}$ and 
	\begin{align}
	& 	\| \hat{V} -V \| \le  |\iX||\iY||\iZ| \sqrt{\| \hat{V}^{XY} -V^{XY} \| } \label{equalitya} \\
	& \DD(\hat{V} \| \hat{V}^{XY}\circ W) \le  \DD(V \| V^{XY}\circ W) \left(1+\sqrt{\| \hat{V}^{XY} -V^{XY} \|}  \right) \label{equalityb} .
	\end{align} 
\end{Lem}
\begin{IEEEproof}
	Define $\hat{V} (x,y,z) = \hat{V} (x,y) \hat{V}(z|x,y)$ by 	
	\begin{align}
	\hat{V}(z|x,y) \triangleq\left\{\begin{array}{lr}
	V(z|x,y)  & \text{ if  } V(x,y) \ge \eta\\
	W(z|x,y) &  \text{ if  } V(x,y) <\eta
	\end{array},
	\right.\label{defVhullam}
	\end{align}
	where $\eta > 0$ will be specified later. Then		
	\begin{align}
	\|\hat{V} - V\|= &\sum_{(x,y):  V(x,y) \ge \eta}\sum_{z \in \iZ} | \hat{V}(x,y) - V(x,y) | V(z|x,y) \notag\\
	&+  \sum_{(x,y):  \hat{V}(x,y) < \eta}\sum_{z \in \iZ} | \hat{V}(x,y) W(z|x,y) - V(x,y,z) |.
	\end{align}	
	Since
	\begin{align} 
	| \hat{V}(x,y) W(z|x,y) - V(x,y,z) | \le | \hat{V}(x,y) - V(x,y) | W(z|x,y) +   V(x,y)|W(z|x,y) - V(z|x,y) |,
	\end{align}	
	it follows that 
	\begin{align}
	\|\hat{V} - V\| \le \| \hat{V}^{XY} -V^{XY} \| + |\{(x,y): V(x,y) < \eta \}| \cdot  \eta \cdot |\iZ|.
	\end{align}	
	If $\eta < (|\iX| |\iY|)^{-1}$ then $|\{(x,y): V(x,y) < \eta \} | \le |\iX| |\iY|-1$, and if also $\| \hat{V}^{XY} -V^{XY} \| \le \eta |\iZ|$ then we get 
	\begin{align}
	\|\hat{V} - V\| \le |\iX||\iY||\iZ| \cdot \eta. \label{harmasvarbecslese}
	\end{align}
	Next, the identity 
	\begin{align}
	\DD(V \| V^{XY}\circ W) = \sum_{(x,y) \in \iX \times \iY} V(x,y)\DD(V(\cdot|x,y)  \| W(\cdot|x,y)) \label{feltdivdef}
	\end{align}
	will be used, that implies, in particular
	\begin{align}
	\DD(V(\cdot|x,y)  \| W(\cdot|x,y)) \le \frac{1}{V(x,y)} \DD(V \| V^{XY}\circ W). \label{feltdivdefkov}
	\end{align}
	By (\ref{feltdivdef}) and its counterpart for $\hat{V}$, (\ref{defVhullam}) and (\ref{feltdivdefkov}) give
	\begin{align}
	&\DD(\hat{V} \| \hat{V}^{XY}\circ W) - \DD(V \| V^{XY}\circ W) \notag\\
	&\le \sum_{(x,y): V(x,y) \ge \eta } \left(\hat{V}(x,y)-V(x,y) \right) \DD(V(\cdot|x,y)  \| W(\cdot|x,y)) \\
	&\le  \| \hat{V}^{XY} -V^{XY} \| \cdot \DD(V \| V^{XY}\circ W) \cdot \eta^{-1}.    \label{divkulfelsobecsles}
	\end{align}
	With the choice $\eta= \sqrt{ \| \hat{V}^{XY} -V^{XY} \|}$, (\ref{harmasvarbecslese}) and (\ref{divkulfelsobecsles}) give (\ref{equalitya}) and (\ref{equalityb}). That choice does meet the conditions in the passage preceding (\ref{harmasvarbecslese}), the first one by assumption, and the second one trivially. \end{IEEEproof}
\begin{Rem}
	A "natural" choice for $\hat{V}$ would be to let $\hat{V}(z|x,y)=V(z|x,y)$ for all $(x,y) \in \iX \times \iY$ with $V(x,y) >0$. This would yield $ \| \hat{V}-V \| = \| \hat{V}^{XY} -V^{XY} \|$, better than (\ref{equalitya}), but does not appear to admit a bound like (\ref{equalityb}) that involves only the variational distance $\| \hat{V}^{XY} -V^{XY} \|$. A bound involving also $W$ could be easily given, but would not suffice for our purposes.  
\end{Rem}
Actually, we will use the following consequence of Lemmas \ref{LemChangeDistribToVP} and \ref{LemContinuityTool}.
\begin{Cor} \label{CorFolyt}
	Given any channel matrix $W$ and distributions  $V \in \iP (\iX \times \iY \times \iZ)$, $\hat{V}^X \in \iP (\iX)$, $\hat{V}^Y \in \iP (\iY)$ with  
	\begin{align}
	& 	\| \hat{V}^X -V^X \| \le \delta, \quad \| \hat{V}^Y -V^Y \| \le \delta , \quad \delta < \frac{1}{2} (|\iX||\iY|)^{-2}, \label{corfolytfeltetelek}
	\end{align} 
	there exists  $\hat{V} \in \iP (\iX \times \iY \times \iZ)$ whose $X$ and $Y$ marginals are the given $\hat{V}^X$, $\hat{V}^Y$, and that satisfies $\| \hat{V}^{XY} -V^{XY} \| \le 2 \delta$ as well as (\ref{equalitya}) and (\ref{equalityb}).
\end{Cor}

\begin{IEEEproof}[Proof of Theorem \ref{ThmExpEgyenletesFolyt}] The notation in the passage preceding Theorem \ref{ThmExpEgyenletesFolyt} is used. Recall that $\delta_n$ and $e_k$ are defined by (\ref{EqDeltan}) and (\ref{Defek}). In particular, $\sum_{k \in [2K]} e_k = K$. 
	
	Consider a sequence of distributions $\vV=(V_1,\dots, V_{2K})$ that attains the minimum in the definition \eqref{DefEnAlphaS} of $E_n^{\alpha} (\vS)$:
	\begin{equation}
	V_k \in \iP_{n,k}, k \in S; \quad E_{\vV}^{\alpha} (\vS) = E_n^{\alpha} (\vS). \label{optimalisvsorozat}
	\end{equation}
	We prove Theorem \ref{ThmExpEgyenletesFolyt} by assigning to $\vV$ a sequence $\hat{\vV}=(\hat{V}_1,\dots,\hat{V}_{2K})$ such that 
	\begin{equation}
	\hat{V}_k \in \iP^*, k \in S; \quad E_{\hat{\vV}}^{\alpha} (\vS) - E_{\vV}^{\alpha} (\vS) \le \gamma_n \label{alternativvsorozattul}
	\end{equation}
	with $\gamma_n$ as in the Theorem. Only the components with index $k \in S$ of $\vV$ and $\hat{\vV}$ have to be considered as those with $k \notin S$ do not affect the value of $E_{\vV}^{\alpha} (\vS)$ resp. $E_{\hat{\vV}}^{\alpha} (\vS)$. Accordingly, in the rest of the proof, always $k \in S$. 
	
	Due to (\ref{EVD}) and the obvious inequalities $|a|^+ - |b|^+ \le |a-b|^+$, $|\sum a_i|^+ \le \sum |a_i|^+$, the difference in  (\ref{alternativvsorozattul}) is bounded above by
	\begin{align}
	& =\sum_{k\in S} e_k \left[\DD(\hat{V}_{k}\|\hat{V}_k^{XY} \circ W ) - \DD(V_{k}\|V_k^{XY }\circ  W) \right]  + \sum_{k\in S} e_k (\I_{\hat{V}_k}^{0} - \I_{V_k}^{0} ) \notag\\
	&\quad +\sum_{k\in S_{1}} e_k |\I_{\hat{V}_k}^{1} - \I_{V_k}^{1}|^{+} + \sum_{k\in S_{2}} e_k |\I_{\hat{V}_k}^{2} - \I_{V_k}^{2} |^+ +\sum_{k\in S_{12}} e_k |\I_{\hat{V}_k}^{12} - \I_{V_k}^{12}|^+. \label{EVDkulonbsegbecsles}
	\end{align}
	With also later purposes in mind, consider the following assignment of distribution $\hat{V}_k \in \iP(\iX \times \iY \times \iZ)$ to $V_k$, $k \in S$, with each $\hat{V}_k$ having prescribed marginals $\hat{V}^X$ and $\hat{V}^Y$ (in (\ref{alternativvsorozattul}), $\hat{V}^X=P^X$, $\hat{V}^Y=P^Y$). Let  $\xi>0$ and $0 < \delta < \frac{1}{8} (|\iX||\iY||\iZ|)^{-2}$ be such that 
	\begin{align}
	& 	\| \hat{V}^X -V_k^X \| \le \delta, \quad \| \hat{V}^Y -V_k^Y \| \le \delta , \quad \text{when } e_k \ge \xi  . \label{corfeltetelek2}
	\end{align} 
	The upper bound to $\delta$, smaller than that in (\ref{corfolytfeltetelek}), will be needed later. 
	
	Set 
	\begin{equation}
	\hat{V}_k (x,y,z)\triangleq \hat{V}^X (x)\hat{V}^Y (y)W(z|x,y) \quad \text{if } e_k < \xi, 
	\end{equation}
	and for indices $k$ with $e_k \ge \xi $ (hence for all $k$ if $\alpha \in [\xi,1-\xi]$) assign $\hat{V}_k$ to $V_k$ according to Corollary \ref{CorFolyt}. 
	
	We will prove the following: For $\{\hat{V}_k, k \in S\}$ as above, the sum (\ref{EVDkulonbsegbecsles}) is bounded above by 
	\begin{equation}
	K \cdot \log (|\iX||\iY||\iZ|) \cdot \xi + E_{\vV}^{\alpha}(\vS) \cdot \sqrt{2 \delta} + 8 \cdot K \cdot |\iX||\iY||\iZ| \cdot \sqrt{2 \delta} \cdot \log \frac{1}{\sqrt{2 \delta}} \label{kulonbsegrekorlat} 
	\end{equation}
	When (\ref{kulonbsegrekorlat}) will be used to establish (\ref{alternativvsorozattul}), the choice of $\delta$ will depend on $\delta_n$ in (\ref{EqDeltan}) but not as directly as the notation may suggest. 
	
	Now, for indices $k$ with $e_k \le \xi$ (if any) trivially  $\DD(\hat{V}_{k}\|\hat{V}_k^{XY} \circ W ) + \I_{\hat{V}_k}^{0} =0$, and $|\I_{\hat{V}_k}^{i} - \I_{V_k}^{i}| \le  \log (|\iX||\iY||\iZ|)$ for $i \in \{1,2,12\}$. Hence, the (less than $K$) terms of (\ref{EVDkulonbsegbecsles}) with such indices $k$ have sum less than $K \cdot \log (|\iX||\iY||\iZ|) \cdot \xi$.
	
	For indices $k$ with $e_k \ge \xi$ we have by Corollary \ref{CorFolyt} 
	\begin{equation}
	\DD(\hat{V}_{k}\|\hat{V}_k^{XY} \circ W ) - \DD(V_{k}\|V_k^{XY }\circ  W) \le \sqrt{2 \delta} \cdot \DD(V_{k}\|V_k^{XY }\circ  W),
	\end{equation}
	hence, the sum of the corresponding terms in (\ref{EVDkulonbsegbecsles}) is bounded by
	\begin{equation}
	\sqrt{2 \delta} \cdot \sum_{k \in S} e_k \DD(V_{k}\|V_k^{XY }\circ  W) \le \sqrt{2 \delta} \cdot E_{\vV}^{\alpha} (\vS). \label{divergenciaboladodobecsles}\end{equation}

	Finally, the sum of the remaining terms of (\ref{EVDkulonbsegbecsles}) can be bounded via employing a bound to entropy differences by variational distances for our purposes, Lemma 2.7 of  \cite{Csiszar} suffices. Since the differences $\I_{\hat{V}_k}^{i} - \I_{V_k}^{i}$, $i \in \{0,1,2,12\}$, can be decomposed into 3 or 4 entropy differences, the cited lemma implies
	\begin{equation}
	|\I_{\hat{V}_k}^{i} - \I_{V_k}^{i}| \le 4 \cdot \| \hat{V}_k -V_k \| \log \frac{|\iX||\iY||\iZ|}{\| \hat{V}_k -V_k \|}, \quad i \in \{0,1,2,12\}, 
	\end{equation}
	if $\| \hat{V}_k -V_k \| \le \nicefrac{1}{2}$ (for $i \in \{0,1,2\}$, the bound has been weakened, for convenience). Using this and the bound to $\| \hat{V}_k -V_k \|$ provided by Corollary \ref{CorFolyt}, it follows that the sum of the corresponding terms of (\ref{EVDkulonbsegbecsles}) is bounded by $8 K (|\iX||\iY||\iZ|)\sqrt{2 \delta} \log \frac{1}{\sqrt{2 \delta}}$; it is here where we have used the assumption $\delta < \frac{1}{8} (|\iX||\iY||\iZ|)^{-2}$. 
	
	Thereby we have established the bound (\ref{kulonbsegrekorlat}) to (\ref{EVDkulonbsegbecsles}). It gives rise to (\ref{alternativvsorozattul}) as follows. If (\ref{optimalisvsorozat}) holds, then (\ref{EqBal4}) imply that $\| V_k^X -P^X \|$ and $\| V_k^Y -P^Y \|$ are less than $\sqrt{2 \ln 2} \sqrt{\frac{\delta_n}{e_k}}$. This means that (\ref{corfeltetelek2}) holds with $\hat{V}^X=P^X$, $\hat{V}^Y=P^Y$ and $\delta\triangleq \sqrt{2 \ln 2} \sqrt{\frac{\delta_n}{e_k}}$ for any choice of $\xi >0$. Note that (\ref{optimalisvsorozat}) implies that
	\begin{equation}
	E_{\vV}^{\alpha}(\vS) \le E_{\vV^0}^{\alpha}(\vS)\le K \log (|\iX||\iY||\iZ|), \label{uniformfelsokorlat}
	\end{equation}
	where $\vV^0\triangleq (P^{XYZ},\dots,P^{XYZ})$. Using this fact, (\ref{kulonbsegrekorlat}) gives a uniform bound $\gamma_n \rightarrow 0$ to $E_{\hat{\vV}}^{\alpha}(\vS) - E_{\vV}^{\alpha}(\vS)$ as required, when $\xi=\xi_n$ is suitable chosen. A suitable choice is $\xi_n = n^{-\frac{1}{5}} \log n$, this gives $\gamma_n$ equal to $C n^{-\frac{1}{5}} \log n$ where $C$ is a constant depending only on $|\iX|$, $|\iY|$, $|\iZ|$ and $K$. \end{IEEEproof}
\begin{Rem}\label{RemNemjoElmeletGyakban}
	This proof falls short of implying that $E_n^{\alpha} (\vS)$ is close to $E^\alpha (\vS)$ for blocklengths occuring in practice. For that, (\ref{alternativvsorozattul}) ought to be established with $\gamma_n$ approaching $0$ substantially faster than guaranteed by the above proof. This problem remains open. 
\end{Rem}

\begin{IEEEproof}[Completion of the proof of Theorem \ref{ThmRovid}] Fix $L \in [2K-2]$ and $j \in \{1,2\}$. Denote $E^{\alpha} (L,j)$ in (\ref{ELJ}) as a function of $\alpha$, $P^X$, $P^Y$, $R_1$, $R_2$, $W$ by $f(\alpha, P^X,P^Y,R_1,R_2)$; recall that the coefficients $\beta_i$, $i \in \{1,2,12\}$ in (\ref{ELJ}) are determined by $L$, $j$ and $\alpha$. Further, denote by $g$ the function of $V_1$, $V_2$, $V_{12}$, $\alpha$, $R_1$, $R_2$, $W$ minimized in (\ref{ELJ}) with respect to $V_1$, $V_2$, $V_{12}$ subject to
	\begin{equation} \label{Eqcon}
	V_i^X = P^X, \quad V_i^Y=P^Y, \quad i \in \{1,2,12\}
	\end{equation}
	Thus, using the identity (\ref{EqMainThmProofDivExchange}),
	\begin{align} \label{Eqg}
	g(V_1,V_2,V_{12}, \alpha, R_1,R_2,W)= \sum_{i \in \{1,2,12\}} \beta_i \left[ \DD( V_{i} \| V_i^{XY} \circ W) + \I_{V_i}^0 \right]  + \left|\sum_{i \in \{1,2,12\}} \beta_i (\I_{V_i}^i -R_i)\right|^+,
	\end{align}
	where $R_{12} \triangleq R_1 + R_2$. 
	
	We have to prove that $f$ is jointly continous in its $6$ variables. This will be done in two steps. 
	
	\emph{Step 1.} We show that for fixed $W$ the values of $f$ at  $\alpha$, $P^X$, $P^Y$, $R_1$, $R_2$ and  $\hat{\alpha}$, $\hat{P}^X$, $\hat{P}^Y$, $\hat{R}_1$, $\hat{R}_2$ with
	\begin{equation} \label{Eqdelta}
	|\alpha-\hat{\alpha}| \le \delta, \text{ } |P^X-\hat{P}^X| \le \delta, \text{ } |P^Y-\hat{P}^Y| \le \delta, \text{ } |R_1-\hat{R}_1| \le \delta, \text{ } |R_2-\hat{R}_2| \le \delta
	\end{equation} 
	do not differ by more than $\varepsilon = \varepsilon (\delta, |\iX|,|\iY|,|\iZ|,K)$, where $\varepsilon \rightarrow 0$ as  $\delta \rightarrow 0$. By symmetry, it suffices to show that 
	\begin{equation} \label{Eqepsilon}
	f(\hat{\alpha}, \hat{P}^X,\hat{P}^Y,\hat{R}_1,\hat{R}_2,W) - f(\alpha, P^X,P^Y,R_1,R_2,W) \le \varepsilon
	\end{equation}
	when (\ref{Eqdelta}) holds. To this end, to $(V_1,V_2,V_{12})$ minimzing $g$ subject to (\ref{Eqcon}) for the given $\alpha$, $P^X$, $P^Y$, $R_1$, $R_2$, $W$ we assign a triple $(\hat{V}_1,\hat{V}_2,\hat{V}_{12})$ satisfying the analogue of (\ref{Eqcon}) with hats, such that
	\begin{equation} \label{Eqge}
	g(\hat{V}_1,\hat{V}_2,\hat{V}_{12},\hat{\alpha},\hat{R}_1,\hat{R}_2,W) - g(V_1,V_2,V_{12},\alpha,R_1,R_2,W) \le \varepsilon
	\end{equation}
	Suitbale distributions $\hat{V}_i, i \in \{1,2,12\}$ will be chosen like $\hat{V}_k, k \in S$ in the proof of Theorem \ref{ThmExpEgyenletesFolyt}: With $\xi >0$ specified later, set 
	\begin{equation}
	\hat{V}_i (x,y,z)\triangleq \hat{P}^X (x)\hat{P}^Y (y)W(z|x,y) \quad \text{if } \beta_i < \xi, 
	\end{equation}
	and if $\beta_i \ge \xi$ then assign $\hat{V}_i$ to $V_i$ according to Corollary \ref{CorFolyt} with $\hat{P}^X$, $\hat{P}^Y$ in the role of $\hat{V}^{X}$, $\hat{V}^{Y}$. Thus, in case $\beta_i \ge \xi$ let $\hat{V}_i$ satisfy
	\begin{equation}
	\hat{V}_i^X = \hat{P}^X, \text{ } \hat{V}_i^Y = \hat{P}^Y, \text{ } \| \hat{V}_i^{XY}-V_i^{XY} \| \le 2 \delta
	\end{equation}
	as well as (\ref{equalitya}) and (\ref{equalityb}) with $V_i$ and $\hat{V}_i$ in the role of $V$ and $\hat{V}$. 
	
	Write the lhs of (\ref{Eqge}) as $\varDelta_1 + \varDelta_2$ where
	\begin{align}
	&\varDelta_1 \triangleq g(\hat{V}_1,\hat{V}_2,\hat{V}_{12},\alpha,\hat{R}_1,\hat{R}_2,W) - g(V_1,V_2,V_{12},\alpha,R_1,R_2,W)  \\
	&\varDelta_2 \triangleq g(\hat{V}_1,\hat{V}_2,\hat{V}_{12},\hat{\alpha},\hat{R}_1,\hat{R}_2,W) - g(\hat{V}_1,\hat{V}_2,\hat{V}_{12},\alpha,\hat{R}_1,\hat{R}_2,W).
	\end{align}
	Here $\varDelta_1$ can be bounded as $E_{\hat{\vV}}^{\alpha} (\vS) - E_{\vV}^{\alpha}(\vS)$ has been in the proof of Theorem \ref{ThmExpEgyenletesFolyt}. It follows from (\ref{Eqg}) as there that
	\begin{align}
	\varDelta_1&\le \sum_{i\in \{1,2,12\}} \beta_i \left[\DD(\hat{V}_{i}\|\hat{V}_i^{XY} \circ W ) - \DD(V_{i}\|V_i^{XY }\circ  W) \right]  + \sum_{i\in \{1,2,12\}} \beta_i (\I_{\hat{V}_i}^{0} - \I_{V_i}^{0} ) \notag\\
	&\quad +\sum_{i\in \{1,2,12\}} \beta_i |\I_{\hat{V}_i}^{i} - \I_{V_i}^{i}|^{+} + \sum_{i\in \{1,2,12\}} \beta_i  |R_i - \hat{R}_i |^+. \label{Eq1b}
	\end{align}
	Disregarding the last sum, the rhs of (\ref{Eq1b}) is of the same form as (\ref{EVDkulonbsegbecsles}), with $i \in \{1,2,12 \}$ and $\beta_i$ playing the role of $k \in S$ and $e_k$. Hence the bound (\ref{kulonbsegrekorlat}) may be employed (its derivation has used that the sum of the coefficients $e_k$ does not exceed $K$, but this holds also for the coefficients $\beta_i$). The role of $E_{\vV}^{\alpha}(\vS)$ in (\ref{kulonbsegrekorlat}) is played by $E^{\alpha}(L,j)=f(\alpha,P^X,P^Y,R_1,R_2)$. It can be bounded by $ K \log (|\iX||\iY||\iZ|)$ as in the proof of Theorem \ref{ThmExpEgyenletesFolyt}. As the rightmost sum in (\ref{Eq1b}) is less than $K\delta$ subject to (\ref{Eqdelta}), it follows that 
	\begin{equation}
	\varDelta_1 \le K \left[ \log (|\iX||\iY||\iZ|) \cdot (\xi + \sqrt{2 \delta}) + 8 \cdot |\iX||\iY||\iZ| \cdot \sqrt{2 \delta} \cdot \log \frac{1}{\sqrt{2 \delta}} +\delta \right], \label{Eq1bb} 
	\end{equation}
	provided that $\delta < \frac{1}{8} (|\iX||\iY||\iZ|)^{-2}$ needed for (\ref{kulonbsegrekorlat}).
	
	Next, $\varDelta_2$ is equal to $0$ if $L$ is odd. If $L$ is even then $\beta_i$ and its counterpart $\hat{\beta}_i$ satisfy $|\hat{\beta}_i - \beta_i|=|\hat{\alpha} - \alpha|$, hence (\ref{Eqg}) and (\ref{Eqdelta}) imply
	\begin{equation} \label{Eq2b}
	\varDelta_2 \le  \delta  \sum_{i \in \{1,2,12\}} \left[ \DD( \hat{V}_{i} \| \hat{V}_i^{XY} \circ W) + \I_{\hat{V}_i}^0  + |\I_{\hat{V}_i}^i -\hat{R}_i |^+ \right].
	\end{equation}
	Here $ \DD( \hat{V}_{i} \| \hat{V}_i^{XY} \circ W)$ is nonzero only if $\beta_i \ge \xi$, and then
	\begin{equation}
	\DD( \hat{V}_{i} \| \hat{V}_i^{XY} \circ W) <  2 \cdot  \DD( V_{i} \| V_i^{XY} \circ W) \le \frac{2}{\xi} E^{\alpha}(L,j) \le \frac{2K \log (|\iX||\iY||\iZ|)}{\xi},
	\end{equation}
	where the first inequality follows from (\ref{equalityb}) and the second one from (\ref{ELJ}). The other terms of the sum in (\ref{Eq2b}) are trivially bounded by constants. 
	
	Choosing $\xi = \sqrt{\delta}$, say, this completes the proof of (\ref{Eqge}), with $\varepsilon = C \sqrt{\delta} \log \frac{1}{\delta}$, where $C$ is a contant depending only on $|\iX|$,$|\iY|$,$|\iZ|$, $K$. 
	
	\emph{Step (ii).} Due to the result of Step(i), it suffices to prove the continuity of $f$ in $W$ when the other variables are fixed. To this, only its lower semicontinuity in $W$ has to be proved, since any function convex and finite valued on a polytop is upper semicontinous there (see \cite{rockafellar-1970a}). 
	
	Clearly, $g$ is lower semicontinous (lsc) in $(V_1,V_2,V_{12},W)$, hence the claim that $f$ is lsc in $W$ is an instance of the following fact. If a function $g(u,v)$ of $(u,v)$ in an Euclidean space is lsc then so is also $f(v)\triangleq \min_{u \in \mathcal{K}} g(u,v)$, if $\mathcal{K}$ is a compact set. To verify it, suppose $v_n \rightarrow v^*$, let $u_n \in \mathcal{K}$ attain $\min_{u \in \mathcal{K}} g(u,v_n)$, and take a sunsequence $v_{n_k}$ such that $f(v_{n_k}) \rightarrow \liminf_{n \rightarrow \infty} f(v_n)$ and $u_{n_k}$ is convergent, say to $u^*$. Then
	\begin{equation}
	\liminf_{n \rightarrow \infty} f(v_n) = \lim_{k \rightarrow \infty} f(v_{n_k})= \lim_{k \rightarrow \infty} g(u_{n_k},v_{n_k})\ge g(u^*,v^*) \ge f(v^*),
	\end{equation}
	thus $f$ is lsc as claimed.
	
	This completes the proof of the continuity assertion of Theorem \ref{ThmRovid}.
\end{IEEEproof}

\section*{Acknowledgment}
 We thank Gerhard Kramer and Arun Padakandla for drawing our attention to \cite{HouSmeePfisterTomasin2006} and \cite{Arunjavaslat}, respectively. We also thank the support of the National Research, Development and Innovation Office – NKFIH K120706 and KH129601.

\end{appendices}

\bibliography{references-arXiv}
\bibliographystyle{IEEEtran}

\end{document}